\documentclass[%
 reprint,
 amsmath,amssymb,
 aps,
prb,
]{revtex4-2}

\usepackage{graphicx}
\usepackage{dcolumn}
\usepackage{bm}
\usepackage{xcolor}
\usepackage{booktabs}
\usepackage{tabu}

\usepackage{xr}
\makeatletter
\newcommand*{\addFileDependency}[1]{
  \typeout{(#1)}
  \@addtofilelist{#1}
  \IfFileExists{#1}{}{\typeout{No file #1.}}
}
\makeatother

\newcommand*{\myexternaldocument}[1]{
    \externaldocument{#1}
    \addFileDependency{#1.tex}
    \addFileDependency{#1.aux}
}

\myexternaldocument{./SI}
\def\br{{\bf r}}
\def\bp{{\bf r}'}

\begin{document}

\preprint{APS/123-QED}

\title{Dynamical downfolding for localized quantum states}

\author{Mariya Romanova}
\author{Guorong Weng}
\author{Arsineh Apelian}
\author{Vojt\v{e}ch Vl\v{c}ek}%
 \email{vlcek@ucsb.edu}
\affiliation{Department of Chemistry and Biochemistry, University of California, Santa Barbara, CA 93106-9510, U.S.A.}%

\date{\today}

\begin{abstract}
We introduce an approach to treat localized correlated electronic states in the otherwise weakly correlated host medium. Here, the environment is dynamically downfolded on the correlated subspace. It is captured via renormalization of one and two quasiparticle interaction terms which are evaluated using many-body perturbation theory. We outline the strategy on how to take the dynamical effects into account by going beyond the static limit approximation. Further, we introduce an efficient stochastic implementation that enables treating the host environment with a large number of electrons at a minimal computational cost. For small explicitly correlated subspace, the dynamical effects are critical. We demonstrate the methodology by reproducing optical excitations in the negatively charged NV center defect in diamond, that are in excellent agreement with experimental results.
\end{abstract}

\maketitle


\section{Introduction}
The ability to predict and computationally tackle electronic excitations is key for guiding the development of new materials in many areas ranging from quantum technologies to ultrafast electronics. In this context, materials hosting strongly coupled electronic states are particularly interesting, but they pose a significant challenge to theory. Despite the great progress in making explicitly correlated computational approaches more affordable,~\cite{wagner2016discovering,tubman2020modern,szalay2012multiconfiguration,li2018fast,blunt2015semi,mejuto2020efficient,orus2019tensor,chan2011density}  
calculations are still limited to problems with a small number of quantum particles. Fortunately, the most important contribution is, in many cases, limited to only a small range of electronic states, i.e., a subspace of the system. A common approach to alleviate the computational cost is to invoke theoretical treatment~\cite{georges2004strongly,libisch2014embedded,cui2019efficient,pham2019periodic,rusakov2018self,ma2021quantum,sheng2022green,dvorak2019dynamical,dvorak2019quantum} in which a small explicitly correlated problem (solvable computationally) is embedded in the remaining portion of the system, which is treated at a more approximate level.

In practice, the calculations have to deal with a number of methodological bottlenecks that can compromise the accuracy of the model. Determining the correlated subspace is not always straightforward and often based on chemical or physical intuition. Further, the strongly interacting states are coupled to the rest of the system~\cite{mejuto2020efficient}, i.e., the remaining (weakly interacting) electrons are influenced by the electronic configuration of the subspace and self-consistent treatment is thus required.~\cite{hampel2020effect,acharya2021importance} Moreover, the extent of the dynamical coupling depends on the size of the explicitly correlated region; in other words: the larger the correlated region is, the more dynamical interactions are treated explicitly and the simpler the coupling to the rest of the system is. The importance of the dynamical renormalization has been recognized~\cite{aryasetiawan2004frequency,aryasetiawan2009downfolded,werner2010dynamical} and the constrained random phase approximation (cRPA) has become a \textit{de facto} standard in accounting for the dynamics of the environment outside of the correlated subspace~\cite{bockstedte2018ab,ma2020quantum,ma2021quantum,muechler2022quantum,sheng2022green}. For technical reasons, a static limit approximation is ubiquitously applied, instead of the fully dynamical description,~~\cite{martin2016interacting,dvorak2019dynamical,dvorak2019quantum} and cRPA calculations are computationally prohibitive for large systems.~\cite{romanova2022stochastic}

Here, we combine an efficient stochastic cRPA (s-cRPA)~\cite{romanova2022stochastic} approach and describe a complementary strategy in which the weakly interacting environment is downfolded on the correlated subspace and the dynamical interactions are fully taken into account. In analogy to the approaches considering individual quasiparticles (QPs), the majority of the system (i.e., the environment) is captured via effective single- and two-QP interactions within the subspace. The dynamical response of the weakly correlated electrons is captured by renormalized interaction terms, whose dynamics compensates for the reduced subspace dimensionality.

We exemplify the approach by reproducing the experimentally measured optical excitations in a single NV$^{-}$ defect center in diamond. The defect consists of a small number of correlated states (requiring an explicitly correlated method), formed by dangling bonds pointing toward the vacancy. This subspace definition is physically and chemically motivated. The remaining host environment is weakly correlated: diamond's fundamental and optical band gaps are well reproduced within many-body perturbation theory (MBPT) already at the level of $G_0W_0$~\cite{Vlcek2018swift,lofaas2011effective,gao2015band} and BSE~\cite{rocca2012solution,leng2016gw}, suggesting that MBPT is sufficient for the electronic dynamics and the related downfolding. Further, we employ stochastic formalism that is applicable to large scale systems. The optical excitations obtained by the stochastic dynamical downfolding approach are in excellent agreement with experimental measurements.
\begin{figure}
    \centering
    \includegraphics[width=3.37in]{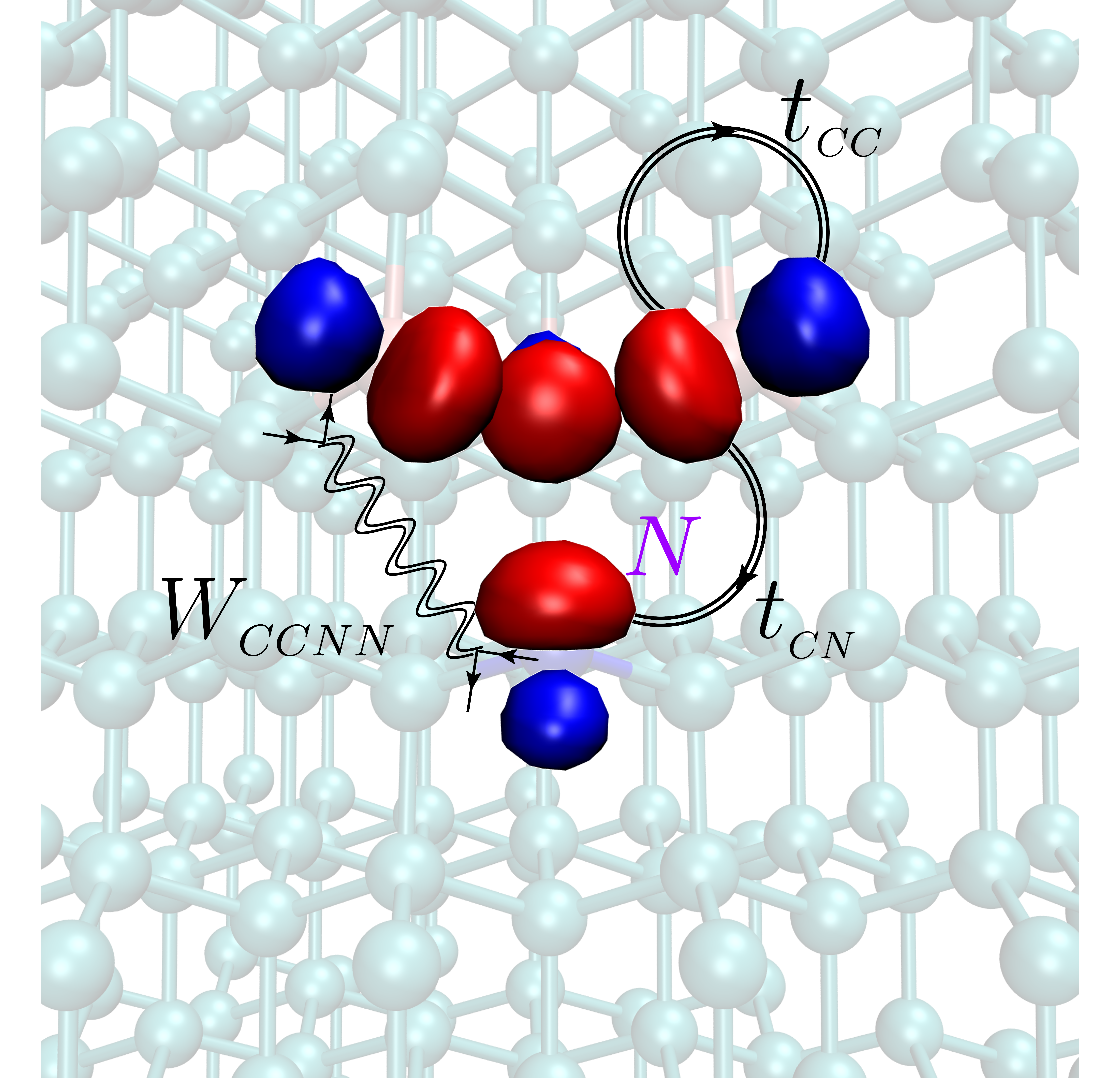}
    \caption{Model representation of the NV$^-$. Three carbon sites are on top and nitrogen site at the bottom. Maximally localized single particle states are shown by isosurface plots: red and blue colors represent positive and negative value of a real valued orbitals. Examples of one- and two-body terms are depicted diagrammatically. The one-body terms ($t$) are obtained from single QP propagators renormalized by the self-energy of the environment. The two-body terms ($W$) contain screened Coulomb interactions. }
    \label{fig:fig_model}
\end{figure}

\section{Theory}

The first step in solving the correlated problem is to project the electronic structure onto a selected subspace of states treated by an explicitly correlated effective Hamiltonian:
\begin{equation}
\begin{split}
\label{eq:hamiltonian}
 \hat{H}=-\sum_{i,j,\sigma}^{}  t_{ij}\hat{c}^{\dagger}_{i\sigma}\hat{c}_{j\sigma}
 + \frac{1}{2}\sum_{ijkl\sigma\sigma'}^{} W_{ijkl} \hat{c}^{\dagger}_{i\sigma} \hat{c}^{\dagger}_{j\sigma'} \hat{c}_{l\sigma'}\hat{c}_{k\sigma},
\end{split}
\end{equation}
where $\hat{c}^{\dagger}_{i\sigma}$ and $\hat{c}_{i,\sigma}$ are creation and annihilation operators in site $i$ with spin $\sigma$. $t_{ij}$ represents the one-body term capturing individual QPs within the subspace. The term is conventionally separated into the on-site, $i=j$, and hopping, $i\neq j$, amplitudes. The $t$ terms are computed as $t_{ij} = \langle i| H^{\rm env}_{qp}|j\rangle$ using a single-QP Hamiltonian, $H^{\rm env}_{qp}$ , containing the downfolded interactions from the environment. The second term,  $W_{ijkl}$, describes two-body (i.e., explicit QP-QP) interactions within the subspace. 

The $t$ and $W$ terms critically depend on the definition of the correlated space and the projection onto it effectively lowers the problem dimensionality. It is exact provided that both one- and two-body terms downfold all the interactions with the remainder of the system.~\cite{martin2016interacting} We consider only electronic degrees of freedom; the downfolded representation thus captures the coupling with the electronic states (i.e., the charge and configuration fluctuations) outside of the subspace treated by Eq.~\eqref{eq:hamiltonian}. In particular, we consider the charge density and exchange-correlation interactions. Both contributions include dynamical induced effects (e.g., polarization) and higher order terms described e.g., in Ref.~\cite{vlcek2019stochastic} and hence renormalize the $t$ and $W$ terms.\cite{martin2016interacting} To the lowest order, the induced interactions include induced charge density fluctuations. In general, they also contain an induced density matrix.~\cite{vlcek2019stochastic,mejuto2022} Note that if the dynamics of the subspace problem is artificially decoupled from the environment (i.e., only static interactions are considered), the exact solution requires a self-consistent re-evaluation of the one- and two-body terms to reflect the induced effects. Such a self-consistency is circumvented in the (dynamical) downfolding approach shown here.

Clearly, the size of the explicitly correlated region and its coupling to the remainder of the system determines the degree of dynamical effects that needs to be captured by $t$ and $W$. In this work, we seek only the minimal configuration space, obtained by localizing the electronic states on the atoms neighboring the defect (Fig.~\ref{fig:fig_model}), and explore the role of dynamical renormalization of individual terms entering Eq.~\eqref{eq:hamiltonian}. This leads to only four, physically motivated, orbitals in which the QPs are renormalized by the environment of weakly correlated electrons in diamond. The practical calculations thus need to address how to determine such terms and how (i.e., at which frequency) they enter the Hamiltonian.

The downfolding derives directly from the single QP equation of motion:
\begin{equation}\label{eq:GF_eom}
   i\partial_{t_1} G (1,2)= \delta(1,2) + h_0(1)G(1,2) - i v(1^+,\bar 3) L(1,\bar 3,2,\bar 3^+),
\end{equation}
where $G$ is the one-body Green's function (GF) defined as $G(1,2)=-i \left \langle N|\mathcal T [\hat c(1) \hat c^\dagger(2)] |N\right\rangle$ with time-ordering operator $\mathcal T$ and $|N\rangle$ representing the $N$-particle many-body ground state. Further, $h_0 = \hat T + \hat V^{\rm ext}$ is the one-particle operator composed of the kinetic and external (ionic) potential terms, $v$ is the bare Coulomb interaction, and $L$ is the two-particle Green's function defined as a time ordered product of two creation and two annihilation operators (i.e., analogously to $G$). We adopted the shorthand notation for space-time coordinates, i.e., $1\equiv r_1, t_1$, and bar indicates a coordinate to be integrated over. In MBPT, the two-body interactions are downfolded by introducing Hartree and exchange-correlation self-energies:
\begin{equation}\label{eq:2bd_factorization}
v(1^+,\bar 3) L(1,\bar 3,2,\bar 3^+) =  \Sigma_H(1,\bar3) G(1,2) + \Sigma_{xc}(1,\bar 3)G(\bar 3,2).
\end{equation}
The Hartree self-energy, $\Sigma_H$,  is the potential due to the density mediated by the instantaneous Coulomb term $\Sigma_H(1,\bar 3) = v(1,\bar 3) G(\bar 3,\bar 3^+)$ and $v(1,2) = (|\br_1 - \br_2|)^{-1}\delta(t_1-t_2)$.  $\Sigma_{xc}$ downfolds all quantum two-body effects not included in $\Sigma_H$ (discussed later and, e.g.,  in Ref.~\cite{martin2016interacting}). The QP dynamics is governed by excitation energies, corresponding to  poles of $G$ in the frequency domain; these are represented as eigenvalues of the effective QP Hamiltonian 
$H^{qp} = \hat T + \hat V^{\rm ext} + \hat \Sigma (\omega)$, where the total self-energy, $\Sigma$, is frequency dependent as a result of Fourier transformation from the time domain. For a particular QP state, $H^{qp}$ is computed as a fixed point equation with  $\Sigma$ evaluated at $\omega$ corresponding to the QP energy.

We now discuss the calculations of $t$ and $W$ terms, which follow a similar route with the important difference, that only the interactions with electrons in the remainder of the systems (i.e., outside of the correlated subspace) are downfolded. Given the constrained description with fixed minimal orbital space, localizing a QP in a particular state $\phi_j$ is associated with charge fluctuation inside the subspace and polarization of the environment. The QPs in the subspace thus interact via a renormalized (screened) Coulomb interaction: 
\begin{equation}\label{eq:Wenv}
    W^{\rm env}(1,2) = \nu(1,2) + \nu(1,\bar3) \chi^{\rm env} (\bar 3,\bar 4) \nu(\bar 4,2),
\end{equation}
where $\chi^{\rm env}$ is the reducible polarizability of the environment due to a potential variation $\delta U$: $\chi^{\rm env} = \delta n^{\rm env}/ \delta U$, where $\delta U$ is in the subspace that is ``external'' to the environment. Within the GF formalism, the two-particle interactions are formally downfolded by substituting $v$ terms by $W^{\rm env}$. 

It is now important to distinguish interactions within the subspace and with the remainder of the system. For a pair or QPs inside the correlated region, we consider explicit interactions mediated by the renormalized $ W^{\rm env}$. Hence, for a particular choice of localized single-QP states $\{\phi\}$, we obtain a dynamical two-body term as:
\begin{equation}
    W_{ijkl}(\omega) = \left\langle \phi_i \phi_j \middle| \hat W^{\rm env}(\omega)\middle|\phi_k \phi_l\right\rangle,  
    \label{eq:Wijkl_def}
\end{equation}
which enters Eq.~\eqref{eq:hamiltonian}. 

For the one-body term, the single-QP dynamics is governed by Eq.~\eqref{eq:GF_eom} with downfolded two-body interactions between electrons inside and outside the correlated subspace. We require, that the same formalism applies (formally) to all states, i.e., the two-body interactions represented by $L$ will be mediated by $W^{\rm env}$. We thus write:
\begin{align}
&W^{\rm env}(1^+,\bar 3)L(1,\bar 3,2,\bar 3^+)\nonumber \\&=  \Sigma^{\rm env}_H(1,\bar3) G(1,2) + \Sigma^{\rm env}_{xc}(1,\bar 3)G(\bar 3,2),\label{eq:renorm_2bdy_factorization}
\end{align}
where $\Sigma^{\rm env}_H (1,\bar 3) = W^{\rm env} (1,\bar 3) G^{\rm env}(\bar 3,\bar 3^+) $ is a dynamical Hartree term which includes the effect of induced charge density in the environment, represented by $G^{\rm env}(1,1^+)$, caused by density fluctuations inside the subspace. Similarly, the exchange correlation term is, in general, $\Sigma^{\rm env}_{xc} (1,2)= G^{\rm env}(1,\bar 3)W^{\rm env}(1^+,\bar 4)\tilde\Gamma^{\rm env}(\bar 3,\bar 4,2)$, where the GF is constructed only from the states of the environment. The vertex $\tilde \Gamma^{\rm env} = \delta (G^{\rm env})^{-1}/\delta U$ is constructed from the response of the environment to the variation of the ``external'' potential $\delta U$. Note that the vertex is reducible, i.e., distinct from the counterpart typically encountered in the Hedin's formalism.\cite{martin2016interacting,mejuto2022,hedin1965new} In practice, this means that an additional screened response needs to be included in the self-energy expansion given that Eq.~\eqref{eq:renorm_2bdy_factorization} substitutes $v$ with $W^{\rm env}$. 

Using the combined self-energy  $\Sigma^{\mathrm{env}} = \Sigma^{\rm env}_{H} + \Sigma^{\rm env}_{xc}$, based on the quantities defined in Eq.~\eqref{eq:renorm_2bdy_factorization}, we obtain the single-QP effective Hamiltonian containing the renormalization effects stemming from the environment:  $H_{qp}^{{\rm env}} = \hat T + \hat V^{\rm ext} + \hat \Sigma^{\mathrm{env}} (\omega)$. As discussed above, $H_{qp}^{{\rm env}}$ directly enters the computation of the one-body terms. In particular, for the onsite term and localized single-QP state $\phi_j$ we obtain: 
\begin{equation}\label{eq:hopping}
   t_{jj} (\omega)=
   \left\langle \phi_j\middle| \hat T + \hat V^{\rm ext} + \hat \Sigma^{\mathrm{env}} (\omega)\middle| \phi_j \right\rangle,
\end{equation}
where $\hat \Sigma^{\mathrm{env}}$ is the self-energy representing the effective interactions among a QP in the subspace and electrons in the environment. The expectation value of $\hat T$ depends only on  $\phi_j$ and hence it is static. Similarly, the external potential ($\hat V^{\rm ext}$) is not frequency dependent, though this may be further generalized if the electron-phonon coupling is taken into account. In this picture, the electron-phonon coupling is represented as a charge induced structural reorganization. As such, in the downfolded treatment, the external potential is independent of the electronic configuration, i.e., static. In principle, this constraint may be lifted in a more generalized case when lattice coupling is taken into account. Hence, both $\hat T$ and $\hat V^{\rm ext}$ capture trivial differences between the subspace sites $j$. Previously, the definition of the $t$ amplitudes was typically limited only to these two terms.\cite{ranjbar2011many, babamoradi2011effect} In contrast, the self-energy introduces non-trivial dynamical effects.

The remaining step to compute the excitation spectrum of the subspace using Eq.~\eqref{eq:hamiltonian} is not straightforward if the renormalized interactions remain functions of frequency, i.e., $t(\omega)$ and $W(\omega)$. Since the representation stems directly from the effective downfolding corresponding to one- and two-body propagators, we consider $t$ and $W$ evaluated at $\omega$ corresponding to the poles of equilibrium $G$ and $L$ in the frequency domain. 

The one-body term is directly linked to the fixed point solution of the QP Hamiltonian, $H_{qp}$, in which $\Sigma$ is evaluated directly at the QP energy, $\omega_{qp}$. The portion of the self-energy that represents the downfolded environment, $\Sigma^{\rm env}$, is also evaluated at  $\omega_{qp}$, as discussed e.g., in Refs.\cite{romanova2020decomposition,sheng2022green}. Since the subspace orbital basis does not diagonalize $H_{qp}$ nor $H_{qp}^{\rm env}$, the off-diagonal (hopping) terms $t_{ij}$, are computed as 
\begin{equation}
t_{ij} = \left.\frac{1}{4} \middle[t_{ij}(\omega_{qp}^i) + t_{ij}(\omega_{qp}^j) + t_{ji}(\omega_{qp}^i) +t_{ji}(\omega_{qp}^j) \right],    
\end{equation}
where $\omega_{qp}^i$ is the $i^{\rm th}$ subspace QP energy. This approach follows the  QP-selfconsistent method~\cite{faleev2004all,bruneval2006effect} that imposes self-adjointness of the Hamiltonian. In our practical calculations (detailed below) we found that the $H_{qp}$ is still strongly diagonally dominant, i.e., $t_{ii}\gg t_{ij}$ and the symmetrized and hermitized form of $t_{ij}$ is justified in these cases \cite{faleev2004all,bruneval2006effect}.
The two-body interactions are renormalized by charge density fluctuations in the remainder of the system, i.e., polarization due to the electron-hole transitions (excitations) in the subspace. In the following, we consider such optical excitations to determine the frequency at which $W_{ijkl}(\omega)$ should be computed (Eq.~\eqref{eq:Wijkl_def}). Note that this is a particular choice; an alternative strategy is to compute the two-body terms from two-particle propagator, e.g., in the $T$-matrix approximation~\cite{martin2016interacting,mejuto2022} in which the two-particle interactions are renormalized by particle-particle scattering (which needs to be restricted to the environment). As we are primarily considering the particle-hole screening in Eq.~\eqref{eq:Wenv}, we employ the Bethe-Salpeter equation for particle-hole propagator $L$. The excitations are defined as the fixed-point solutions for a two-QP Hamiltonian, $H_{2qp}$,~\cite{onida2002electronic} which for a particular set of subspace states is: $ \langle ij| H_{2qp} (\omega) |kl\rangle  = \Delta_{ij}\delta_{ik}\delta_{jl} + \mathcal K_{ijkl}(\omega)$. Here, $\Delta_{ij} = \omega_{qp}^i - \omega_{qp}^j$ is the difference between two QP energies, corresponding to two independent single-QP excitations. Further,  $\mathcal K$ is a general dynamical interaction kernel that couples the two QPs. We consider that the two-body interactions inside the subspace are mediated by the screened Coulomb interaction at the random phase approximation (RPA) level ($\mathcal K_{ijkl} = W_{ijkl}^{\rm env}$). This choice is motivated by two reasons: for the description of individual QPs, we resort to the $GW$ approximation (see below) in which we employ RPA; further, this approach directly utilizes the quantity of interest, i.e., $W_{ijkl}(\omega)$ from Eq.~\eqref{eq:Wijkl_def}. The resulting particle-hole excitations are the fixed point solutions of $ H_{2qp} (\omega)$ evaluated at the excitation energy $\omega_{2qp}$. In the following, we assume that the dynamically renormalized two-body interactions are screened (by the polarization of the environment) in the same way as the electron excitations in the subspace. Hence the two-body interactions
from Eq.~\eqref{eq:Wijkl_def} enter the Hamiltonian (Eq.~\eqref{eq:hamiltonian}) as $W_{ijkl}(\omega_{2qp})$.

\subsection{Stochastic formalism}
We now comment on the practical and efficient implementation of the one and two-body terms. The key ingredient is the renormalized two-body interaction, Eq.~\eqref{eq:Wenv} which directly enters the evaluation of $W_{ijkl}$ and $t$ (Eqs.~\eqref{eq:Wijkl_def} and ~\eqref{eq:hopping}). In practice, we employ a stochastic evaluation of the QP-QP and single-QP terms, i.e., we sample the action of the screened interaction: instead of computing the environment polarizability, $\chi^{\rm env}$, we repeatedly compute the induced density $\delta n^{\rm env} = \chi^{\rm env} \delta U$ constructed from random vectors projected on the occupied portion of the single particle Hilbert space.\cite{neuhauser2014breaking,Vlcek2018swift} The stochastic occupied states, $|\eta\rangle$, are projected such that they are a part of the weakly correlated environment: $|\tilde \eta\rangle = (1-P_\phi) |\eta\rangle$, where $|\eta\rangle$ spans the entire occupied subspace. Here $P_\phi$ is the projection operator explicitly formed from the states of the correlated subspace, $\{\phi\}$:
\begin{equation}\label{projectorphi_occ}
    P_\phi = \sum_{k\in \left\{\phi\right\} } f_k \left| \phi_k\middle\rangle \middle \langle \phi_k \right|,
\end{equation}
where $f_k$ is the occupation of the $k^{\rm th}$ state. The variation of the charge density, $\delta n$ is induced by electrons in particular states of the correlated region $\{\phi\}$, which create a perturbing potential at $t=0$. The density is constructed from random vectors as $n(\br,t) = \lim_{N_\zeta \to \infty} \frac{1}{N_\zeta} \sum_{j=1}^{N_\zeta} |\tilde\eta(\br, t)|^2 $. In practice, the response is computed separately for each perturbation (i.e., for each set of perturbing states $\{\phi\}$); yet the overall cost is significantly reduced compared to the conventional (deterministic) method as only a few sampling vectors $|\tilde \eta\rangle$ are necessary to converge the one- and two-body terms.  This is discussed in SI Sec.~\ref{sec:conv} in more detail. The time evolution of $|\tilde \eta\rangle$ states is evaluated in the RPA, in which only the (mean field) ground state Hamiltonian explicitly depends on time only through the Hartree potential.~\cite{neuhauser2014breaking, Vlcek2018swift, vlcek2017stochastic, vlcek2019stochastic, romanova2022stochastic}. 

For the two-body terms, we conventionally separate the bare and polarization contributions $W_{ijkl}(\omega) = W_{ijkl}^\mathrm{b} + W_{ijkl}^\mathrm{p}(\omega)$, where the former is directly computed as the static part of Eq.~\eqref{eq:Wijkl_def}, and the latter is obtained by Fourier transformation from its time-dependent form: 
\begin{equation}
    \langle \phi_i \phi_j|\tilde W^{\mathrm{p}} (t)| \phi_k \phi_l\rangle  =  \int \phi^*_i(\br)\phi_j(\br) \tilde u(\br,t) d\br.
\label{eq:wp}    
\end{equation}
We use a time ordered induced potential $u$, computed from its retarded counterpart: $\tilde u^r(1) = v(1,\bar 2) \delta n^{\rm  env}(\bar 2)$, where the density fluctuation is a response to a perturbation from the correlated subspace. The time-evolved states $|\tilde \eta\rangle$ are repeatedly projected using $P_\phi$ at each time step to exclude the dynamics of the correlated subspace. This constitutes the stochastic density constrained RPA (s-cRPA) introduced in Ref.~\cite{romanova2022stochastic}. 

Finally, the contribution to the renormalized Hartree self-energy, $\Sigma_H^{\rm env}$ is similar in nature to Eq.~\eqref{eq:wp}, as it also contains the dynamically screened $W^{\rm env}$ term computed via stochastic sampling. However, instead of treating only a set of correlated states $\{\phi\}$, it applies to the interaction between the density of the environment and a individual single-QP densities, represented by $|\phi_j(\br)|^2$. As such it is part of the one-body $t$ term.

In practical calculations, further approximations need to be introduced for the sake of tractability. As argued earlier, the $G_0W_0$ approximation, which captures the correlation effects via induced charge density fluctuations, provides an excellent choice.~\cite{Vlcek2018swift,lofaas2011effective,gao2015band} For evaluating the $\Sigma_{xc}^{\rm env}$ in the $t$ term, we resort to a formulation analogous to $G_0W_0$ as well. In practice, this means that $\tilde\Gamma^{\rm env}(1,2,3) \approx \delta(1,2)\delta(1,3)$. As a result, this form neglects the higher order screening in $\Sigma^{\rm env}_{xc}$, that is inherently part of $\tilde \Gamma$. While this step is an ad-hoc approximation, it still corresponds to the leading-order term capturing the environment exchange-correlation self-energy. 

In this approximation, our calculations employ a portion of the $G_0W_0$ environment exchange-correlation self-energy, which depends only on an underlying mean-field Hamiltonian used to generate the starting point. Analogously to the two-body terms and the Hartree self-energy, $\Sigma_{xc}^{\rm env}$ is efficiently evaluated using a real-time propagation of stochastic vectors (sampling the fluctuations in the environment induced by addition of a particle or hole to the subspace). This step follows the stochastic decomposition scheme detailed in Ref.~\cite{romanova2020decomposition}. In the space-time representation, $G_0^{\rm env}(\br, \bp, t) = \{ \tilde \xi^*(t,\bp)\tilde \zeta(\br)\}$, where $|\tilde\zeta\rangle$ is obtained from a random state sampling the entire single-particle Hilbert space, $|\zeta\rangle$, by projection $|\tilde\zeta\rangle = (1- P_\phi) |\zeta\rangle$. Further, $| \xi(t)\rangle$ is a random state in either occupied or unoccupied subspace of the environment states (obtained by filtering) which is propagated backward or forward in time due to time ordering applied to holes and particles.\cite{neuhauser2014breaking, Vlcek2018swift, vlcek2017stochastic, vlcek2019stochastic} Again, only its portion $|\tilde \xi\rangle$, which is orthogonal to the correlated subspace, contributes to $G^{\rm env}$ and it is prepared as $|\tilde\xi\rangle = P_\phi |\xi\rangle$. Finally, the time evolution is governed by the $U_0$ which depends only on the ground state non-interacting Hamiltonian and $\left| \tilde \xi(t)\right\rangle = U_0(t) \left| \tilde \xi\right\rangle$. The practical expression for the xc self-energy for a state $\phi_j$ in Eq.~\eqref{eq:hopping} thus becomes (in the time domain): $\langle \phi_j | \Sigma (t) | \phi_j \rangle = \langle \phi_j  \tilde\xi(t) | \hat W^{\rm env}(t) |\tilde \zeta \phi_j\rangle $.

\subsection{Implementation}

The following calculation workflow was employed in this work. The atomic relaxations of the NV$^-$ center defect in 3D periodic diamond supercells were performed in the QuantumESPRESSO code\cite{QE2017}. Next, the starting-point DFT calculations were performed with our real-space DFT implementation. The maximally localized functions were obtained by PMWannier2.0 code based on sequential optimization.~\cite{gwen_wannier,weng2022reduced} The localized orbitals were centered on four sites: nitrogen atom and three nearest to vacancy carbon atoms (Fig.~\ref{fig:fig_model}). This is a minimal model of the NV-center that is commonly used~\cite{choi2012mechanism,muechler2022quantum,ranjbar2011many} to describe its low-lying excited states. Further, the model parameters were computed in the localized basis. Screening was computed with the s-cRPA method, that was implemented within a development version of the StochasticGW code.\cite{neuhauser2014breaking, Vlcek2018swift, vlcek2017stochastic, romanova2020decomposition,romanova2022stochastic} In our calculations, we employed in total 3,200 samplings for the one and two body terms; a detailed study of the convergence of the stochastic errors and computational details are in the SI Sec.~\ref{sec:conv} and Sec.~\ref{sec:methods}.

\section{Results}

We first analyze the QP band gap of pristine diamond and compare it to one of the defect supercells. Our stochastic $G_0W_0$ calculations of a 4,096-atom diamond supercell provide a QP band gap of the $5.6$~eV. This is consistent with our previous calculations.~\cite{Vlcek2018swift} Further, the $G_0W_0$ calculations of a 511-atom diamond supercell containing the NV$^{-1}$ defect provide a band gap of $5.3$~eV. This is less than 0.1~eV away from our result for a 512-atom pristine diamond supercell,~\cite{Vlcek2018swift} i.e., the band-gap of the host material is practically unchanged due to the presence of the defect. Both results are in good agreement with the experimental value of $5.5$~eV\cite{nebel2003thin}. Note that there is a significant error cancellation which makes the $G_0W_0$ gap coincide with the experimental zero-phonon gap. In the same vein, it is reasonable to assume that the dynamical downfolding based on $G_0W_0$ and RPA will also perform well and match experimental zero-phonon optical transitions.

\begin{figure*}
    \centering
    \includegraphics[width=6.74in]{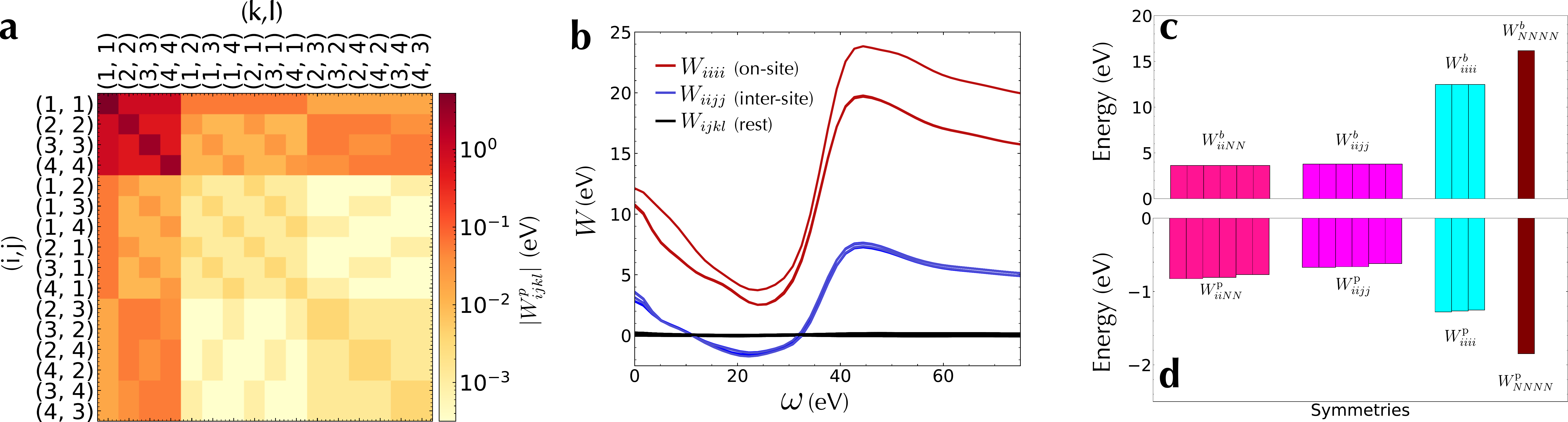}
    \caption{  a. Color map of absolute values of the screening part of all 256 interaction parameters estimated at the static limit $W^\mathrm{p}_{ijkl}(\omega=0)$. b. Frequency dependence of the total bare plus screened interaction for 3200 stochastic samples. Interaction terms are highlighted by color according to the physical meaning. Red represents the on-site Coulomb interaction terms, blue - inter-site, black - all remaining interaction terms. Stochastic errors are smaller than the line thickness. c. Bare Hubbard Coulomb on-site and inter-site parameters $W^\mathrm{b}_{iiii,iijj}$. d. Screening part of the Hubbard Coulomb parameters after retaining the hermitian part of the interactions $W^\mathrm{p}_{iiii,iijj}(\omega=0)$. Computed with 3200 stochastic samples in the 511-atom cell. In c. and d. parameters obeying the same symmetry operations are highlighted with the same color.}
    \label{fig:fig_panel}
\end{figure*}

We now turn to the analysis of the QP energies of the defect defined by the maximally localized states. Treating the defect states at the $G_0W_0$ level leads to a small difference between the individual sites (Fig.~\ref{fig:fig_model}), i.e., they are energetically similar with a difference of merely 0.51~eV for the orbitals located on nitrogen and carbon atoms. Note that at this point all electrons, including those in the correlated subspace, are described by the $G_0W_0$ self-energy. 

The situation changes, for the individual one-body (single QP) terms as they are based on screened interactions and also contain only a portion of the exchange-correlation self-energy. The difference between the $t_{ii}$ terms for the N and C sites becomes significantly larger (12.7~eV). This is because the response of the environment to the localized QPs is not balanced by the induced fluctuation of the particles in the localized states (i.e., only the response of the environment is included in the $t$ terms). This value should be contrasted with the estimate of the $t$ terms neglecting the $\Sigma^{\rm env}$ term (19.9~eV), which represents a simplified version of the onsite applied in previous studies. In this case, it thus seems that the environmental contributions are significantly renormalizing the trivial (kinetic and local external potential) contribution to the differences between the subspace sites. In the Hubbard model, the differences between the onsite $t$ terms are critical to determining the triplet-triplet transition (c.f., SI Sec.~\ref{sec:NVmodel}).
When inspecting the matrix of $t$ elements, including the intersite hopping ($t_{ij}$), we notice that the maximally localized basis is strongly diagonally dominant and the $t_{ij}$ are on average just 6\% of the $t_{ii}$ terms (0.74~eV on average). Further, the dependence of the $t$  terms on $\omega$ is relatively weak and slowly varying. If the dynamical effects are neglected in the Hartree term (i.e., taking the static limit of $\Sigma_H^{\rm env }$), we recover the difference of 15.2~eV between the onsite single-QP terms for N and C orbitals. Here, the renormalization decreases by roughly 35\%, while much of the remaining difference stems from the dynamical contribution of $\Sigma_{xc}^{\rm env}$.

Next we analyze the results for the two-body interactions, $W_{ijkl}$. Figure~\ref{fig:fig_panel}~a shows the values of interaction parameters at the static limit $W^\mathrm{p}(\omega=0)$ on a color map. By ordering the $(i,j)$ and $(k,l)$ indices, we group together the the dominant on-site and inter-site density-density interactions that appear in the left top corner, i.e.,  $W^\mathrm{p}_{iiii}$ and $W^\mathrm{p}_{iijj}$. The remaining terms do not exceed $70$~meV (both bare and screened); this is true for all frequencies as we show in Fig.~\ref{fig:fig_panel}~b (black curves). The convergence of the four dominant parameters with samplings as a function of supercell size is provided in Fig.~\ref{fig:fig4} of the Sec.~\ref{sec:conv} in SI. In the rest of this work, we exclude all the terms with a magnitude of $<70$~meV. 

Note that the minimal model preserves the $C_{3v}$ symmetry, which is also satisfied by the underlying mean-field Hamiltonian. However, the statistical sampling is associated with finite stochastic errors and the corresponding interaction symmetry is restored a posteriori. In particular, the bare component, $W^{\rm b}_{ijkl}$, naturally preserves symmetry under the permutation of indices $ijkl \leftrightarrow klij$. In general, this is not satisfied for stochastically sampled polarization part, $W^{\rm p}_{ijkl}$, (cf.~Fig.~\ref{fig:fig_160_sampl} in SI) which would translate to level splitting (of otherwise degenerate excitations) in the optical spectrum of the NV$^-$. 

To recover the expected behavior we enforce the individual symmetries; in practice, this is analogous to the Hermitization of the QP Hamiltonian as in Refs.~\cite{faleev2004all,bruneval2006effect,romanova2022stochastic}:
\begin{equation}
    W_{ijkl}^\mathrm{p}(\omega) = \frac{1}{4} [W_{ijkl}^\mathrm{p}(\omega) + W_{klij}^\mathrm{p*}(\omega) + W_{klij}^\mathrm{p}(\omega) + W_{ijkl}^\mathrm{p*}(\omega)]
    \label{hermitization}
\end{equation}
Fig.~\ref{fig:fig_panel}d demonstrates the converged result of $W_{ijkl}(\omega)$ with 3,200 samplings. There however remains a finite difference within each group of elements on the order of $20$~meV, attributed to numerical artifacts. This small symmetry breaking leads to errors much lower than the desired resolution of our model $\sim54$~meV, which is below the magnitude of the $W_{ijkl}$ discussed above and shown by the black line in Fig.~\ref{fig:fig_panel}~b. We report the splitting in Table~\ref{tab:table1} in SI for each supercell size. In the rest of this work, we average the screened counterpart of the interaction parameters within each symmetry group (depicted in Fig.~\ref{fig:fig_panel}d for the four cases considered).

In contrast to the $t$ terms, the frequency dependence of the two-body interactions is more pronounced and suggests a larger degree of renormalization. The values of $W_{ijkl}$ are determined at excitation energies obtained from the fixed point solution to $H_{2qp}$; for the optical transition between two C sites, we obtain a value of 2.3~eV while between N and C site a value of 2.6~eV. The energy significantly increases, if excitation is confined to be in between two states on a single site (6.3 and 7.7~eV on the C and N sites). Note that these excitations define the frequency $\omega_{2qp}$, not the optical spectrum we seek. Compared to the bare interactions, the dynamically screened $W_{iiii}(\omega_{2qp})$ are decreased by as much as 52\% for N and 50\% for the C sites. The intersite density-density is lower by 38\% between two C sites, and by 29\% for those between N and C sites. This significant change can be deduced from the steep dependence of $W(\omega)$ curves in Fig.~\ref{fig:fig_panel}~b. As mentioned above, the remaining terms have small magnitudes and are neglected. Taking the static limit of $W(\omega)$ leads to a much weaker screening, especially for the ``on-site'' ($W_{iiii}$) terms, which are reduced by 23\% and 20\%  for N and C sites. In contrast, the intersite terms ($W_{iijj}$) are reduced by 22\% between N and C and by 18\% between two C sites. Regardless of the approach, considering the dynamical effects away from the static limit leads to a reduction of the two-body interaction strength compared to considering the limit $\omega\to0$ (see SI Sec.~\ref{sec:NVmodel}).

\begin{figure}
    \centering
    \includegraphics[width=3.37in]{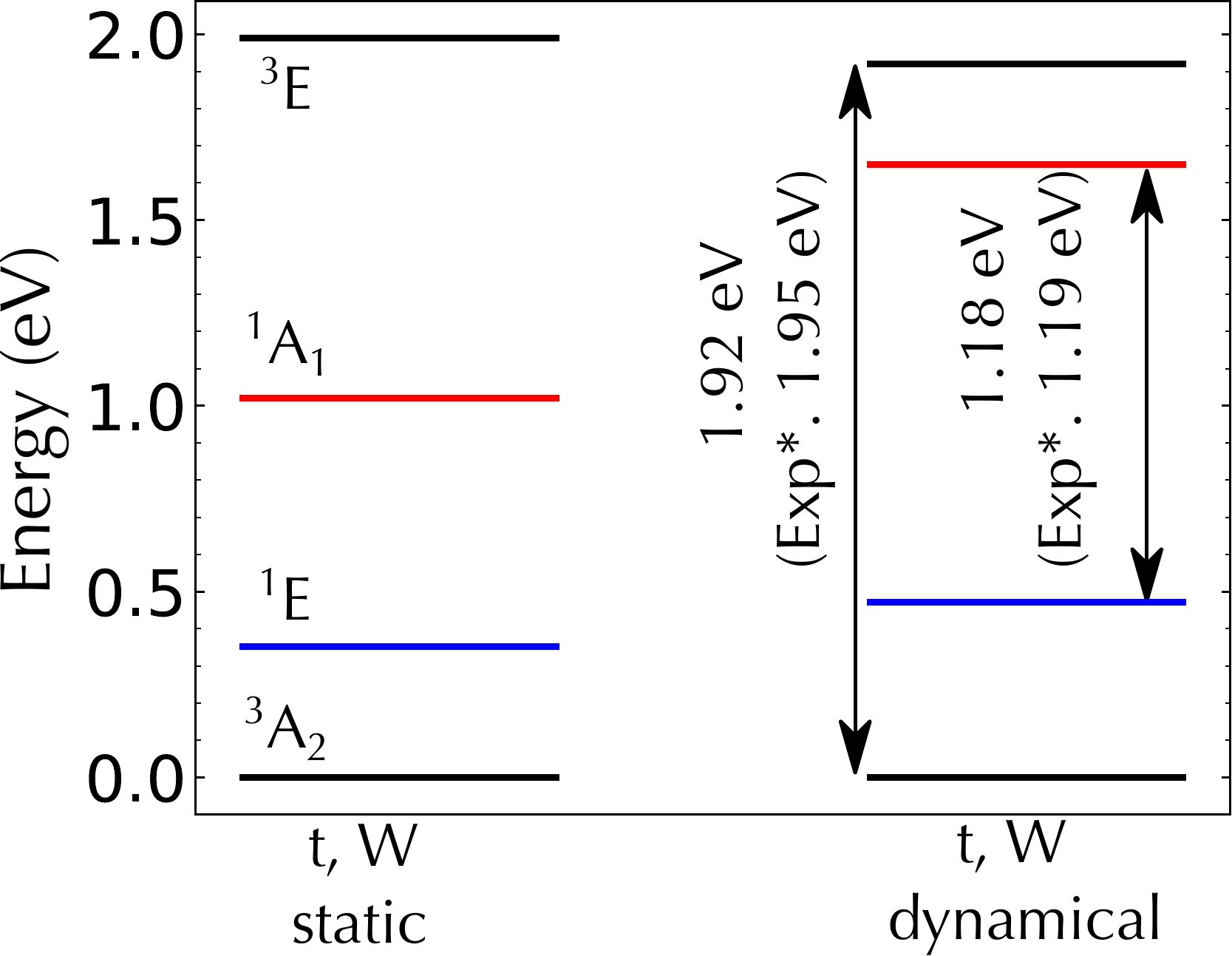}
    \caption{Comparison of the excited states of the NV$^-$ center in 511 atom supercell where the electron-electron interaction parameters were computed in this work with a. t and W screened within stochastic s-cRPA at static limit $\omega=0$ b. both t and W are dynamically renormalized; Number of stochastic samples $N_{{\tilde\eta}}=3200$. (*) Note, experimental values of zero phonon lines are presented on the plot, while our results are computed for vertical excitations.}
    \label{fig:fig_hub_levels}
\end{figure}

In the final step, we study the excitation energies obtained by exact diagonalization of Eq.~\eqref{eq:hamiltonian} within the minimal subspace defined by the localized orbitals. We consider the fully dynamically renormalized case and compare it to the corresponding static limits. In the latter, the one-body terms contain the contribution of $\Sigma^{\rm env}_{xc}$ computed at the single-QP excitation energy (in analogy to the $G_0W_0$ approximation), but the dynamics of the Hartree term is neglected as we take $
\lim_{\omega\to0}\Sigma_H^{\rm env}(\omega)$. Similarly, the two-body interaction in the static limit is  $\lim_{\omega\to0}W_{ijkl}(\omega)$. Figure~\ref{fig:fig_hub_levels} shows the comparison of the two sets of results. 

The triplet-triplet vertical transition, $^3E\,\leftrightarrow \,^3A_2$, is not substantially different for the two limiting cases. Based on the analysis of the simple Hubbard model for the NV$^-$ center (see SI Sec.~\ref{sec:NVmodel}), we expect that the triplet transitions are largely determined by the differences between the onsite one-body terms $t$. While the dynamical effects change them by roughly 20\%, the resulting optical excitation spacing is insensitive to such a change. The static and the dynamically screened limits yield values of 1.98 and 1.92~eV which are in excellent agreement with the experimental zero-phonon excitation of 1.95~eV.

Our result compares well to previous theoretical works that employed embedding methods despite major differences in the theory. For instance, values of $2.02$~eV and $2.05$~eV of the triplet-triplet transition were obtained in Refs.~\cite{bockstedte2018ab,muechler2022quantum} using the quantum defect embedding theory. These approaches employ a hopping term described at the DFT level and the two-body interaction parameters computed with cRPA in the static limit. A double counting error, i.e., a spurious inclusion of (a portion of) the correlated subspace in the calculation for the environment, arises in embedding approaches.~\cite{bockstedte2018ab,muechler2022quantum,sheng2022green} (It is circumvented in this work by separation of the self-energy). This is especially problematic for embedding within DFT;  in Refs.~\cite{bockstedte2018ab,muechler2022quantum} an approximate Hartree-Fock double counting correction scheme was employed. More recent work~\cite{sheng2022green} formulated an exact double counting correction for $G_0W_0$, combined with cRPA treatment of the two-body interactions at the static limit. We surmise that this $\omega \to 0$ formulation translates into a much larger subspace size requirement (indeed, the converged subspace required 12 orbitals, 22 electrons). Ultimately, the converged value for the triplet-triplet transition in Ref.~\cite{sheng2022green} was estimated to be $2.15$~eV, i.e., larger than our result, but in better agreement with absorption data .~\cite{davies1976optical}

The dynamical effects turn out more critical for the singlet-singlet $^1A_1\,\leftrightarrow\, ^1E$ excitations, which are more sensitive to the relative magnitude of the $W$ terms. As mentioned above, the screening significantly decreases the magnitude of $W_{ijkl}$. This is illustrated in the SI and also discussed in more detail below. We see that in the static case, the singlet-singlet transition is $0.7$~eV, which is lower than $1.19$~eV the zero-phonon line observed experimentally. These results are in good agreement with other published cRPA results in the static limit. Specifically, a value of $0.8$~eV was obtained in Refs.~\cite{muechler2022quantum, sheng2022green}, though this can be improved if beyond cRPA response is used~\cite{ma2021quantum}.

In contrast, our result for the fully dynamical renormalization of single-singlet transition is $1.18$~eV (at s-cRPA level) in excellent agreement with the experiment. This is largely due to the reduction of the two-body terms when the dynamics of the density fluctuations is fully taken into account. As noted above, we made a particular choice of $H_{2qp}$ while a different formulation may change the results as the QP-QP interactions would be evaluated at distinct values of $\omega_{2qp}$. Nevertheless, the interaction term is reduced at a \textit{finite frequency} compared to the static limit, and this leads to an improved singlet-singlet transition energy (See Fig.~\ref{fig:fig_U_vary} in SI Sec.~\ref{sec:NVmodel}). This step clearly illustrates that going beyond the typical static limit improves the results appreciably.

\section{Conclusions}

The efficiency of the stochastic dynamical downfolding approach suggests that it is a practical and reliable methodology for treating localized correlated phenomena (such as quantum defects and beyond).  In practice, it captures the problem via a chosen correlated subspace while the host material is represented via the dynamical effect entering each interaction term. Formally, the complexity of the dynamical terms compensates for the small size of the subspace; for weakly correlated host systems (which are accurately treated via MBPT) our strategy offers a great computational advantage. 

Indeed, the results for the negatively charged NV center in bulk diamond practically demonstrate that the dynamically downfolded representation is in excellent agreement with the experimental excited energies of the defect. We believe that this is a jumping-off point for future practical simulations of electronic excitations in localized quantum states.

\begin{acknowledgments}
This material is based upon work supported by the U.S. Department of Energy, Office of Science, Office of Advanced Scientific Computing Research, Scientific Discovery through Advanced Computing (SciDAC) program under Award Number DE-SC0022198. This research used resources of the National Energy Research
Scientific Computing Center, a DOE Office of Science User Facility
supported by the Office of Science of the U.S. Department of Energy
under Contract No. DE-AC02-05CH11231 using NERSC award
BES-ERCAP0020089.

The calculations were also performed as part of the XSEDE\cite{Towns_2014} computational Project No.~TG-CHE180051. Use was made of computational facilities purchased with funds from the National Science Foundation (CNS-1725797) and administered by the Center for Scientific Computing (CSC). The CSC is supported by the California NanoSystems Institute and the Materials Research Science and Engineering Center (MRSEC; NSF DMR-1720256) at UC Santa Barbara. A.A. was supported by the National Science Foundation Graduate Research Fellowship under Grant No. (2139319).

\end{acknowledgments}

\bibliography{biblio}

\begin{thebibliography}{50}%
\makeatletter
\providecommand \@ifxundefined [1]{%
 \@ifx{#1\undefined}
}%
\providecommand \@ifnum [1]{%
 \ifnum #1\expandafter \@firstoftwo
 \else \expandafter \@secondoftwo
 \fi
}%
\providecommand \@ifx [1]{%
 \ifx #1\expandafter \@firstoftwo
 \else \expandafter \@secondoftwo
 \fi
}%
\providecommand \natexlab [1]{#1}%
\providecommand \enquote  [1]{``#1''}%
\providecommand \bibnamefont  [1]{#1}%
\providecommand \bibfnamefont [1]{#1}%
\providecommand \citenamefont [1]{#1}%
\providecommand \href@noop [0]{\@secondoftwo}%
\providecommand \href [0]{\begingroup \@sanitize@url \@href}%
\providecommand \@href[1]{\@@startlink{#1}\@@href}%
\providecommand \@@href[1]{\endgroup#1\@@endlink}%
\providecommand \@sanitize@url [0]{\catcode `\\12\catcode `\$12\catcode
  `\&12\catcode `\#12\catcode `\^12\catcode `\_12\catcode `\%12\relax}%
\providecommand \@@startlink[1]{}%
\providecommand \@@endlink[0]{}%
\providecommand \url  [0]{\begingroup\@sanitize@url \@url }%
\providecommand \@url [1]{\endgroup\@href {#1}{\urlprefix }}%
\providecommand \urlprefix  [0]{URL }%
\providecommand \Eprint [0]{\href }%
\providecommand \doibase [0]{https://doi.org/}%
\providecommand \selectlanguage [0]{\@gobble}%
\providecommand \bibinfo  [0]{\@secondoftwo}%
\providecommand \bibfield  [0]{\@secondoftwo}%
\providecommand \translation [1]{[#1]}%
\providecommand \BibitemOpen [0]{}%
\providecommand \bibitemStop [0]{}%
\providecommand \bibitemNoStop [0]{.\EOS\space}%
\providecommand \EOS [0]{\spacefactor3000\relax}%
\providecommand \BibitemShut  [1]{\csname bibitem#1\endcsname}%
\let\auto@bib@innerbib\@empty
\bibitem [{\citenamefont {Wagner}\ and\ \citenamefont
  {Ceperley}(2016)}]{wagner2016discovering}%
  \BibitemOpen
  \bibfield  {author} {\bibinfo {author} {\bibfnamefont {L.~K.}\ \bibnamefont
  {Wagner}}\ and\ \bibinfo {author} {\bibfnamefont {D.~M.}\ \bibnamefont
  {Ceperley}},\ }\bibfield  {title} {\bibinfo {title} {Discovering correlated
  fermions using quantum monte carlo},\ }\href@noop {} {\bibfield  {journal}
  {\bibinfo  {journal} {Reports on Progress in Physics}\ }\textbf {\bibinfo
  {volume} {79}},\ \bibinfo {pages} {094501} (\bibinfo {year}
  {2016})}\BibitemShut {NoStop}%
\bibitem [{\citenamefont {Tubman}\ \emph {et~al.}(2020)\citenamefont {Tubman},
  \citenamefont {Freeman}, \citenamefont {Levine}, \citenamefont {Hait},
  \citenamefont {Head-Gordon},\ and\ \citenamefont
  {Whaley}}]{tubman2020modern}%
  \BibitemOpen
  \bibfield  {author} {\bibinfo {author} {\bibfnamefont {N.~M.}\ \bibnamefont
  {Tubman}}, \bibinfo {author} {\bibfnamefont {C.~D.}\ \bibnamefont {Freeman}},
  \bibinfo {author} {\bibfnamefont {D.~S.}\ \bibnamefont {Levine}}, \bibinfo
  {author} {\bibfnamefont {D.}~\bibnamefont {Hait}}, \bibinfo {author}
  {\bibfnamefont {M.}~\bibnamefont {Head-Gordon}},\ and\ \bibinfo {author}
  {\bibfnamefont {K.~B.}\ \bibnamefont {Whaley}},\ }\bibfield  {title}
  {\bibinfo {title} {Modern approaches to exact diagonalization and selected
  configuration interaction with the adaptive sampling ci method},\ }\href@noop
  {} {\bibfield  {journal} {\bibinfo  {journal} {Journal of chemical theory and
  computation}\ }\textbf {\bibinfo {volume} {16}},\ \bibinfo {pages} {2139}
  (\bibinfo {year} {2020})}\BibitemShut {NoStop}%
\bibitem [{\citenamefont {Szalay}\ \emph {et~al.}(2012)\citenamefont {Szalay},
  \citenamefont {Muller}, \citenamefont {Gidofalvi}, \citenamefont {Lischka},\
  and\ \citenamefont {Shepard}}]{szalay2012multiconfiguration}%
  \BibitemOpen
  \bibfield  {author} {\bibinfo {author} {\bibfnamefont {P.~G.}\ \bibnamefont
  {Szalay}}, \bibinfo {author} {\bibfnamefont {T.}~\bibnamefont {Muller}},
  \bibinfo {author} {\bibfnamefont {G.}~\bibnamefont {Gidofalvi}}, \bibinfo
  {author} {\bibfnamefont {H.}~\bibnamefont {Lischka}},\ and\ \bibinfo {author}
  {\bibfnamefont {R.}~\bibnamefont {Shepard}},\ }\bibfield  {title} {\bibinfo
  {title} {Multiconfiguration self-consistent field and multireference
  configuration interaction methods and applications},\ }\href@noop {}
  {\bibfield  {journal} {\bibinfo  {journal} {Chemical reviews}\ }\textbf
  {\bibinfo {volume} {112}},\ \bibinfo {pages} {108} (\bibinfo {year}
  {2012})}\BibitemShut {NoStop}%
\bibitem [{\citenamefont {Li}\ \emph {et~al.}(2018)\citenamefont {Li},
  \citenamefont {Otten}, \citenamefont {Holmes}, \citenamefont {Sharma},\ and\
  \citenamefont {Umrigar}}]{li2018fast}%
  \BibitemOpen
  \bibfield  {author} {\bibinfo {author} {\bibfnamefont {J.}~\bibnamefont
  {Li}}, \bibinfo {author} {\bibfnamefont {M.}~\bibnamefont {Otten}}, \bibinfo
  {author} {\bibfnamefont {A.~A.}\ \bibnamefont {Holmes}}, \bibinfo {author}
  {\bibfnamefont {S.}~\bibnamefont {Sharma}},\ and\ \bibinfo {author}
  {\bibfnamefont {C.~J.}\ \bibnamefont {Umrigar}},\ }\bibfield  {title}
  {\bibinfo {title} {Fast semistochastic heat-bath configuration interaction},\
  }\href@noop {} {\bibfield  {journal} {\bibinfo  {journal} {The Journal of
  chemical physics}\ }\textbf {\bibinfo {volume} {149}},\ \bibinfo {pages}
  {214110} (\bibinfo {year} {2018})}\BibitemShut {NoStop}%
\bibitem [{\citenamefont {Blunt}\ \emph {et~al.}(2015)\citenamefont {Blunt},
  \citenamefont {Smart}, \citenamefont {Kersten}, \citenamefont {Spencer},
  \citenamefont {Booth},\ and\ \citenamefont {Alavi}}]{blunt2015semi}%
  \BibitemOpen
  \bibfield  {author} {\bibinfo {author} {\bibfnamefont {N.}~\bibnamefont
  {Blunt}}, \bibinfo {author} {\bibfnamefont {S.~D.}\ \bibnamefont {Smart}},
  \bibinfo {author} {\bibfnamefont {J.}~\bibnamefont {Kersten}}, \bibinfo
  {author} {\bibfnamefont {J.}~\bibnamefont {Spencer}}, \bibinfo {author}
  {\bibfnamefont {G.~H.}\ \bibnamefont {Booth}},\ and\ \bibinfo {author}
  {\bibfnamefont {A.}~\bibnamefont {Alavi}},\ }\bibfield  {title} {\bibinfo
  {title} {Semi-stochastic full configuration interaction quantum monte carlo:
  Developments and application},\ }\href@noop {} {\bibfield  {journal}
  {\bibinfo  {journal} {The Journal of chemical physics}\ }\textbf {\bibinfo
  {volume} {142}},\ \bibinfo {pages} {184107} (\bibinfo {year}
  {2015})}\BibitemShut {NoStop}%
\bibitem [{\citenamefont {Mejuto-Zaera}\ \emph {et~al.}(2020)\citenamefont
  {Mejuto-Zaera}, \citenamefont {Zepeda-N{\'u}{\~n}ez}, \citenamefont
  {Lindsey}, \citenamefont {Tubman}, \citenamefont {Whaley},\ and\
  \citenamefont {Lin}}]{mejuto2020efficient}%
  \BibitemOpen
  \bibfield  {author} {\bibinfo {author} {\bibfnamefont {C.}~\bibnamefont
  {Mejuto-Zaera}}, \bibinfo {author} {\bibfnamefont {L.}~\bibnamefont
  {Zepeda-N{\'u}{\~n}ez}}, \bibinfo {author} {\bibfnamefont {M.}~\bibnamefont
  {Lindsey}}, \bibinfo {author} {\bibfnamefont {N.}~\bibnamefont {Tubman}},
  \bibinfo {author} {\bibfnamefont {B.}~\bibnamefont {Whaley}},\ and\ \bibinfo
  {author} {\bibfnamefont {L.}~\bibnamefont {Lin}},\ }\bibfield  {title}
  {\bibinfo {title} {Efficient hybridization fitting for dynamical mean-field
  theory via semi-definite relaxation},\ }\href@noop {} {\bibfield  {journal}
  {\bibinfo  {journal} {Physical Review B}\ }\textbf {\bibinfo {volume}
  {101}},\ \bibinfo {pages} {035143} (\bibinfo {year} {2020})}\BibitemShut
  {NoStop}%
\bibitem [{\citenamefont {Or{\'u}s}(2019)}]{orus2019tensor}%
  \BibitemOpen
  \bibfield  {author} {\bibinfo {author} {\bibfnamefont {R.}~\bibnamefont
  {Or{\'u}s}},\ }\bibfield  {title} {\bibinfo {title} {Tensor networks for
  complex quantum systems},\ }\href@noop {} {\bibfield  {journal} {\bibinfo
  {journal} {Nature Reviews Physics}\ }\textbf {\bibinfo {volume} {1}},\
  \bibinfo {pages} {538} (\bibinfo {year} {2019})}\BibitemShut {NoStop}%
\bibitem [{\citenamefont {Chan}\ and\ \citenamefont
  {Sharma}(2011)}]{chan2011density}%
  \BibitemOpen
  \bibfield  {author} {\bibinfo {author} {\bibfnamefont {G.~K.-L.}\
  \bibnamefont {Chan}}\ and\ \bibinfo {author} {\bibfnamefont {S.}~\bibnamefont
  {Sharma}},\ }\bibfield  {title} {\bibinfo {title} {The density matrix
  renormalization group in quantum chemistry},\ }\href@noop {} {\bibfield
  {journal} {\bibinfo  {journal} {Annual review of physical chemistry}\
  }\textbf {\bibinfo {volume} {62}},\ \bibinfo {pages} {465} (\bibinfo {year}
  {2011})}\BibitemShut {NoStop}%
\bibitem [{\citenamefont {Georges}(2004)}]{georges2004strongly}%
  \BibitemOpen
  \bibfield  {author} {\bibinfo {author} {\bibfnamefont {A.}~\bibnamefont
  {Georges}},\ }\bibfield  {title} {\bibinfo {title} {Strongly correlated
  electron materials: Dynamical mean-field theory and electronic structure},\
  }in\ \href@noop {} {\emph {\bibinfo {booktitle} {AIP Conference
  Proceedings}}},\ Vol.\ \bibinfo {volume} {715}\ (\bibinfo {organization}
  {American Institute of Physics},\ \bibinfo {year} {2004})\ pp.\ \bibinfo
  {pages} {3--74}\BibitemShut {NoStop}%
\bibitem [{\citenamefont {Libisch}\ \emph {et~al.}(2014)\citenamefont
  {Libisch}, \citenamefont {Huang},\ and\ \citenamefont
  {Carter}}]{libisch2014embedded}%
  \BibitemOpen
  \bibfield  {author} {\bibinfo {author} {\bibfnamefont {F.}~\bibnamefont
  {Libisch}}, \bibinfo {author} {\bibfnamefont {C.}~\bibnamefont {Huang}},\
  and\ \bibinfo {author} {\bibfnamefont {E.~A.}\ \bibnamefont {Carter}},\
  }\bibfield  {title} {\bibinfo {title} {Embedded correlated wavefunction
  schemes: Theory and applications},\ }\href@noop {} {\bibfield  {journal}
  {\bibinfo  {journal} {Accounts of chemical research}\ }\textbf {\bibinfo
  {volume} {47}},\ \bibinfo {pages} {2768} (\bibinfo {year}
  {2014})}\BibitemShut {NoStop}%
\bibitem [{\citenamefont {Cui}\ \emph {et~al.}(2019)\citenamefont {Cui},
  \citenamefont {Zhu},\ and\ \citenamefont {Chan}}]{cui2019efficient}%
  \BibitemOpen
  \bibfield  {author} {\bibinfo {author} {\bibfnamefont {Z.-H.}\ \bibnamefont
  {Cui}}, \bibinfo {author} {\bibfnamefont {T.}~\bibnamefont {Zhu}},\ and\
  \bibinfo {author} {\bibfnamefont {G.~K.-L.}\ \bibnamefont {Chan}},\
  }\bibfield  {title} {\bibinfo {title} {Efficient implementation of ab initio
  quantum embedding in periodic systems: Density matrix embedding theory},\
  }\href@noop {} {\bibfield  {journal} {\bibinfo  {journal} {Journal of
  Chemical Theory and Computation}\ }\textbf {\bibinfo {volume} {16}},\
  \bibinfo {pages} {119} (\bibinfo {year} {2019})}\BibitemShut {NoStop}%
\bibitem [{\citenamefont {Pham}\ \emph {et~al.}(2019)\citenamefont {Pham},
  \citenamefont {Hermes},\ and\ \citenamefont {Gagliardi}}]{pham2019periodic}%
  \BibitemOpen
  \bibfield  {author} {\bibinfo {author} {\bibfnamefont {H.~Q.}\ \bibnamefont
  {Pham}}, \bibinfo {author} {\bibfnamefont {M.~R.}\ \bibnamefont {Hermes}},\
  and\ \bibinfo {author} {\bibfnamefont {L.}~\bibnamefont {Gagliardi}},\
  }\bibfield  {title} {\bibinfo {title} {Periodic electronic structure
  calculations with the density matrix embedding theory},\ }\href@noop {}
  {\bibfield  {journal} {\bibinfo  {journal} {Journal of Chemical Theory and
  Computation}\ }\textbf {\bibinfo {volume} {16}},\ \bibinfo {pages} {130}
  (\bibinfo {year} {2019})}\BibitemShut {NoStop}%
\bibitem [{\citenamefont {Rusakov}\ \emph {et~al.}(2018)\citenamefont
  {Rusakov}, \citenamefont {Iskakov}, \citenamefont {Tran},\ and\ \citenamefont
  {Zgid}}]{rusakov2018self}%
  \BibitemOpen
  \bibfield  {author} {\bibinfo {author} {\bibfnamefont {A.~A.}\ \bibnamefont
  {Rusakov}}, \bibinfo {author} {\bibfnamefont {S.}~\bibnamefont {Iskakov}},
  \bibinfo {author} {\bibfnamefont {L.~N.}\ \bibnamefont {Tran}},\ and\
  \bibinfo {author} {\bibfnamefont {D.}~\bibnamefont {Zgid}},\ }\bibfield
  {title} {\bibinfo {title} {Self-energy embedding theory (seet) for periodic
  systems},\ }\href@noop {} {\bibfield  {journal} {\bibinfo  {journal} {Journal
  of chemical theory and computation}\ }\textbf {\bibinfo {volume} {15}},\
  \bibinfo {pages} {229} (\bibinfo {year} {2018})}\BibitemShut {NoStop}%
\bibitem [{\citenamefont {Ma}\ \emph {et~al.}(2021)\citenamefont {Ma},
  \citenamefont {Sheng}, \citenamefont {Govoni},\ and\ \citenamefont
  {Galli}}]{ma2021quantum}%
  \BibitemOpen
  \bibfield  {author} {\bibinfo {author} {\bibfnamefont {H.}~\bibnamefont
  {Ma}}, \bibinfo {author} {\bibfnamefont {N.}~\bibnamefont {Sheng}}, \bibinfo
  {author} {\bibfnamefont {M.}~\bibnamefont {Govoni}},\ and\ \bibinfo {author}
  {\bibfnamefont {G.}~\bibnamefont {Galli}},\ }\bibfield  {title} {\bibinfo
  {title} {Quantum embedding theory for strongly correlated states in
  materials},\ }\href@noop {} {\bibfield  {journal} {\bibinfo  {journal}
  {Journal of Chemical Theory and Computation}\ }\textbf {\bibinfo {volume}
  {17}},\ \bibinfo {pages} {2116} (\bibinfo {year} {2021})}\BibitemShut
  {NoStop}%
\bibitem [{\citenamefont {Sheng}\ \emph {et~al.}(2022)\citenamefont {Sheng},
  \citenamefont {Vorwerk}, \citenamefont {Govoni},\ and\ \citenamefont
  {Galli}}]{sheng2022green}%
  \BibitemOpen
  \bibfield  {author} {\bibinfo {author} {\bibfnamefont {N.}~\bibnamefont
  {Sheng}}, \bibinfo {author} {\bibfnamefont {C.}~\bibnamefont {Vorwerk}},
  \bibinfo {author} {\bibfnamefont {M.}~\bibnamefont {Govoni}},\ and\ \bibinfo
  {author} {\bibfnamefont {G.}~\bibnamefont {Galli}},\ }\bibfield  {title}
  {\bibinfo {title} {Green’s function formulation of quantum defect embedding
  theory},\ }\href@noop {} {\bibfield  {journal} {\bibinfo  {journal} {Journal
  of Chemical Theory and Computation}\ } (\bibinfo {year} {2022})}\BibitemShut
  {NoStop}%
\bibitem [{\citenamefont {Dvorak}\ and\ \citenamefont
  {Rinke}(2019)}]{dvorak2019dynamical}%
  \BibitemOpen
  \bibfield  {author} {\bibinfo {author} {\bibfnamefont {M.}~\bibnamefont
  {Dvorak}}\ and\ \bibinfo {author} {\bibfnamefont {P.}~\bibnamefont {Rinke}},\
  }\bibfield  {title} {\bibinfo {title} {Dynamical configuration interaction:
  Quantum embedding that combines wave functions and green's functions},\
  }\href@noop {} {\bibfield  {journal} {\bibinfo  {journal} {Physical Review
  B}\ }\textbf {\bibinfo {volume} {99}},\ \bibinfo {pages} {115134} (\bibinfo
  {year} {2019})}\BibitemShut {NoStop}%
\bibitem [{\citenamefont {Dvorak}\ \emph {et~al.}(2019)\citenamefont {Dvorak},
  \citenamefont {Golze},\ and\ \citenamefont {Rinke}}]{dvorak2019quantum}%
  \BibitemOpen
  \bibfield  {author} {\bibinfo {author} {\bibfnamefont {M.}~\bibnamefont
  {Dvorak}}, \bibinfo {author} {\bibfnamefont {D.}~\bibnamefont {Golze}},\ and\
  \bibinfo {author} {\bibfnamefont {P.}~\bibnamefont {Rinke}},\ }\bibfield
  {title} {\bibinfo {title} {Quantum embedding theory in the screened coulomb
  interaction: Combining configuration interaction with g w/bse},\ }\href@noop
  {} {\bibfield  {journal} {\bibinfo  {journal} {Physical Review Materials}\
  }\textbf {\bibinfo {volume} {3}},\ \bibinfo {pages} {070801} (\bibinfo {year}
  {2019})}\BibitemShut {NoStop}%
\bibitem [{\citenamefont {Hampel}\ \emph {et~al.}(2020)\citenamefont {Hampel},
  \citenamefont {Beck},\ and\ \citenamefont {Ederer}}]{hampel2020effect}%
  \BibitemOpen
  \bibfield  {author} {\bibinfo {author} {\bibfnamefont {A.}~\bibnamefont
  {Hampel}}, \bibinfo {author} {\bibfnamefont {S.}~\bibnamefont {Beck}},\ and\
  \bibinfo {author} {\bibfnamefont {C.}~\bibnamefont {Ederer}},\ }\bibfield
  {title} {\bibinfo {title} {Effect of charge self-consistency in dft+ dmft
  calculations for complex transition metal oxides},\ }\href@noop {} {\bibfield
   {journal} {\bibinfo  {journal} {Physical Review Research}\ }\textbf
  {\bibinfo {volume} {2}},\ \bibinfo {pages} {033088} (\bibinfo {year}
  {2020})}\BibitemShut {NoStop}%
\bibitem [{\citenamefont {Acharya}\ \emph {et~al.}(2021)\citenamefont
  {Acharya}, \citenamefont {Pashov}, \citenamefont {Rudenko}, \citenamefont
  {R{\"o}sner}, \citenamefont {van Schilfgaarde},\ and\ \citenamefont
  {Katsnelson}}]{acharya2021importance}%
  \BibitemOpen
  \bibfield  {author} {\bibinfo {author} {\bibfnamefont {S.}~\bibnamefont
  {Acharya}}, \bibinfo {author} {\bibfnamefont {D.}~\bibnamefont {Pashov}},
  \bibinfo {author} {\bibfnamefont {A.~N.}\ \bibnamefont {Rudenko}}, \bibinfo
  {author} {\bibfnamefont {M.}~\bibnamefont {R{\"o}sner}}, \bibinfo {author}
  {\bibfnamefont {M.}~\bibnamefont {van Schilfgaarde}},\ and\ \bibinfo {author}
  {\bibfnamefont {M.~I.}\ \bibnamefont {Katsnelson}},\ }\bibfield  {title}
  {\bibinfo {title} {Importance of charge self-consistency in first-principles
  description of strongly correlated systems},\ }\href@noop {} {\bibfield
  {journal} {\bibinfo  {journal} {npj Computational Materials}\ }\textbf
  {\bibinfo {volume} {7}},\ \bibinfo {pages} {1} (\bibinfo {year}
  {2021})}\BibitemShut {NoStop}%
\bibitem [{\citenamefont {Aryasetiawan}\ \emph {et~al.}(2004)\citenamefont
  {Aryasetiawan}, \citenamefont {Imada}, \citenamefont {Georges}, \citenamefont
  {Kotliar}, \citenamefont {Biermann},\ and\ \citenamefont
  {Lichtenstein}}]{aryasetiawan2004frequency}%
  \BibitemOpen
  \bibfield  {author} {\bibinfo {author} {\bibfnamefont {F.}~\bibnamefont
  {Aryasetiawan}}, \bibinfo {author} {\bibfnamefont {M.}~\bibnamefont {Imada}},
  \bibinfo {author} {\bibfnamefont {A.}~\bibnamefont {Georges}}, \bibinfo
  {author} {\bibfnamefont {G.}~\bibnamefont {Kotliar}}, \bibinfo {author}
  {\bibfnamefont {S.}~\bibnamefont {Biermann}},\ and\ \bibinfo {author}
  {\bibfnamefont {A.}~\bibnamefont {Lichtenstein}},\ }\bibfield  {title}
  {\bibinfo {title} {Frequency-dependent local interactions and low-energy
  effective models from electronic structure calculations},\ }\href@noop {}
  {\bibfield  {journal} {\bibinfo  {journal} {Physical Review B}\ }\textbf
  {\bibinfo {volume} {70}},\ \bibinfo {pages} {195104} (\bibinfo {year}
  {2004})}\BibitemShut {NoStop}%
\bibitem [{\citenamefont {Aryasetiawan}\ \emph {et~al.}(2009)\citenamefont
  {Aryasetiawan}, \citenamefont {Tomczak}, \citenamefont {Miyake},\ and\
  \citenamefont {Sakuma}}]{aryasetiawan2009downfolded}%
  \BibitemOpen
  \bibfield  {author} {\bibinfo {author} {\bibfnamefont {F.}~\bibnamefont
  {Aryasetiawan}}, \bibinfo {author} {\bibfnamefont {J.~M.}\ \bibnamefont
  {Tomczak}}, \bibinfo {author} {\bibfnamefont {T.}~\bibnamefont {Miyake}},\
  and\ \bibinfo {author} {\bibfnamefont {R.}~\bibnamefont {Sakuma}},\
  }\bibfield  {title} {\bibinfo {title} {Downfolded self-energy of
  many-electron systems},\ }\href@noop {} {\bibfield  {journal} {\bibinfo
  {journal} {Physical review letters}\ }\textbf {\bibinfo {volume} {102}},\
  \bibinfo {pages} {176402} (\bibinfo {year} {2009})}\BibitemShut {NoStop}%
\bibitem [{\citenamefont {Werner}\ and\ \citenamefont
  {Millis}(2010)}]{werner2010dynamical}%
  \BibitemOpen
  \bibfield  {author} {\bibinfo {author} {\bibfnamefont {P.}~\bibnamefont
  {Werner}}\ and\ \bibinfo {author} {\bibfnamefont {A.~J.}\ \bibnamefont
  {Millis}},\ }\bibfield  {title} {\bibinfo {title} {Dynamical screening in
  correlated electron materials},\ }\href@noop {} {\bibfield  {journal}
  {\bibinfo  {journal} {Physical review letters}\ }\textbf {\bibinfo {volume}
  {104}},\ \bibinfo {pages} {146401} (\bibinfo {year} {2010})}\BibitemShut
  {NoStop}%
\bibitem [{\citenamefont {Bockstedte}\ \emph {et~al.}(2018)\citenamefont
  {Bockstedte}, \citenamefont {Sch{\"u}tz}, \citenamefont {Garratt},
  \citenamefont {Iv{\'a}dy},\ and\ \citenamefont {Gali}}]{bockstedte2018ab}%
  \BibitemOpen
  \bibfield  {author} {\bibinfo {author} {\bibfnamefont {M.}~\bibnamefont
  {Bockstedte}}, \bibinfo {author} {\bibfnamefont {F.}~\bibnamefont
  {Sch{\"u}tz}}, \bibinfo {author} {\bibfnamefont {T.}~\bibnamefont {Garratt}},
  \bibinfo {author} {\bibfnamefont {V.}~\bibnamefont {Iv{\'a}dy}},\ and\
  \bibinfo {author} {\bibfnamefont {A.}~\bibnamefont {Gali}},\ }\bibfield
  {title} {\bibinfo {title} {Ab initio description of highly correlated states
  in defects for realizing quantum bits},\ }\href@noop {} {\bibfield  {journal}
  {\bibinfo  {journal} {npj Quantum Materials}\ }\textbf {\bibinfo {volume}
  {3}},\ \bibinfo {pages} {1} (\bibinfo {year} {2018})}\BibitemShut {NoStop}%
\bibitem [{\citenamefont {Ma}\ \emph {et~al.}(2020)\citenamefont {Ma},
  \citenamefont {Govoni},\ and\ \citenamefont {Galli}}]{ma2020quantum}%
  \BibitemOpen
  \bibfield  {author} {\bibinfo {author} {\bibfnamefont {H.}~\bibnamefont
  {Ma}}, \bibinfo {author} {\bibfnamefont {M.}~\bibnamefont {Govoni}},\ and\
  \bibinfo {author} {\bibfnamefont {G.}~\bibnamefont {Galli}},\ }\bibfield
  {title} {\bibinfo {title} {Quantum simulations of materials on near-term
  quantum computers},\ }\href@noop {} {\bibfield  {journal} {\bibinfo
  {journal} {npj Computational Materials}\ }\textbf {\bibinfo {volume} {6}},\
  \bibinfo {pages} {1} (\bibinfo {year} {2020})}\BibitemShut {NoStop}%
\bibitem [{\citenamefont {Muechler}\ \emph {et~al.}(2022)\citenamefont
  {Muechler}, \citenamefont {Badrtdinov}, \citenamefont {Hampel}, \citenamefont
  {Cano}, \citenamefont {R{\"o}sner},\ and\ \citenamefont
  {Dreyer}}]{muechler2022quantum}%
  \BibitemOpen
  \bibfield  {author} {\bibinfo {author} {\bibfnamefont {L.}~\bibnamefont
  {Muechler}}, \bibinfo {author} {\bibfnamefont {D.~I.}\ \bibnamefont
  {Badrtdinov}}, \bibinfo {author} {\bibfnamefont {A.}~\bibnamefont {Hampel}},
  \bibinfo {author} {\bibfnamefont {J.}~\bibnamefont {Cano}}, \bibinfo {author}
  {\bibfnamefont {M.}~\bibnamefont {R{\"o}sner}},\ and\ \bibinfo {author}
  {\bibfnamefont {C.~E.}\ \bibnamefont {Dreyer}},\ }\bibfield  {title}
  {\bibinfo {title} {Quantum embedding methods for correlated excited states of
  point defects: Case studies and challenges},\ }\href@noop {} {\bibfield
  {journal} {\bibinfo  {journal} {Physical Review B}\ }\textbf {\bibinfo
  {volume} {105}},\ \bibinfo {pages} {235104} (\bibinfo {year}
  {2022})}\BibitemShut {NoStop}%
\bibitem [{\citenamefont {Martin}\ \emph {et~al.}(2016)\citenamefont {Martin},
  \citenamefont {Reining},\ and\ \citenamefont
  {Ceperley}}]{martin2016interacting}%
  \BibitemOpen
  \bibfield  {author} {\bibinfo {author} {\bibfnamefont {R.~M.}\ \bibnamefont
  {Martin}}, \bibinfo {author} {\bibfnamefont {L.}~\bibnamefont {Reining}},\
  and\ \bibinfo {author} {\bibfnamefont {D.~M.}\ \bibnamefont {Ceperley}},\
  }\href@noop {} {\emph {\bibinfo {title} {Interacting Electrons}}}\ (\bibinfo
  {publisher} {Cambridge University Press},\ \bibinfo {year}
  {2016})\BibitemShut {NoStop}%
\bibitem [{\citenamefont {Romanova}\ and\ \citenamefont
  {Vl{\v{c}}ek}(2022)}]{romanova2022stochastic}%
  \BibitemOpen
  \bibfield  {author} {\bibinfo {author} {\bibfnamefont {M.}~\bibnamefont
  {Romanova}}\ and\ \bibinfo {author} {\bibfnamefont {V.}~\bibnamefont
  {Vl{\v{c}}ek}},\ }\bibfield  {title} {\bibinfo {title} {Stochastic many-body
  calculations of moir{\'e} states in twisted bilayer graphene at high
  pressures},\ }\href@noop {} {\bibfield  {journal} {\bibinfo  {journal} {npj
  Computational Materials}\ }\textbf {\bibinfo {volume} {8}},\ \bibinfo {pages}
  {1} (\bibinfo {year} {2022})}\BibitemShut {NoStop}%
\bibitem [{\citenamefont {Vl\v{c}ek}\ \emph {et~al.}(2018)\citenamefont
  {Vl\v{c}ek}, \citenamefont {Li}, \citenamefont {Baer}, \citenamefont
  {Rabani},\ and\ \citenamefont {Neuhauser}}]{Vlcek2018swift}%
  \BibitemOpen
  \bibfield  {author} {\bibinfo {author} {\bibfnamefont {V.}~\bibnamefont
  {Vl\v{c}ek}}, \bibinfo {author} {\bibfnamefont {W.}~\bibnamefont {Li}},
  \bibinfo {author} {\bibfnamefont {R.}~\bibnamefont {Baer}}, \bibinfo {author}
  {\bibfnamefont {E.}~\bibnamefont {Rabani}},\ and\ \bibinfo {author}
  {\bibfnamefont {D.}~\bibnamefont {Neuhauser}},\ }\bibfield  {title} {\bibinfo
  {title} {Swift {GW} beyond 10,000 electrons using sparse stochastic
  compression},\ }\href {https://doi.org/10.1103/PhysRevB.98.075107} {\bibfield
   {journal} {\bibinfo  {journal} {Phys. Rev. B}\ }\textbf {\bibinfo {volume}
  {98}},\ \bibinfo {pages} {075107} (\bibinfo {year} {2018})}\BibitemShut
  {NoStop}%
\bibitem [{\citenamefont {L{\"o}f{\aa}s}\ \emph {et~al.}(2011)\citenamefont
  {L{\"o}f{\aa}s}, \citenamefont {Grigoriev}, \citenamefont {Isberg},\ and\
  \citenamefont {Ahuja}}]{lofaas2011effective}%
  \BibitemOpen
  \bibfield  {author} {\bibinfo {author} {\bibfnamefont {H.}~\bibnamefont
  {L{\"o}f{\aa}s}}, \bibinfo {author} {\bibfnamefont {A.}~\bibnamefont
  {Grigoriev}}, \bibinfo {author} {\bibfnamefont {J.}~\bibnamefont {Isberg}},\
  and\ \bibinfo {author} {\bibfnamefont {R.}~\bibnamefont {Ahuja}},\ }\bibfield
   {title} {\bibinfo {title} {Effective masses and electronic structure of
  diamond including electron correlation effects in first principles
  calculations using the gw-approximation},\ }\href@noop {} {\bibfield
  {journal} {\bibinfo  {journal} {AIP Advances}\ }\textbf {\bibinfo {volume}
  {1}},\ \bibinfo {pages} {032139} (\bibinfo {year} {2011})}\BibitemShut
  {NoStop}%
\bibitem [{\citenamefont {Gao}(2015)}]{gao2015band}%
  \BibitemOpen
  \bibfield  {author} {\bibinfo {author} {\bibfnamefont {S.-P.}\ \bibnamefont
  {Gao}},\ }\bibfield  {title} {\bibinfo {title} {Band gaps and dielectric
  functions of cubic and hexagonal diamond polytypes calculated by many-body
  perturbation theory},\ }\href@noop {} {\bibfield  {journal} {\bibinfo
  {journal} {physica status solidi (b)}\ }\textbf {\bibinfo {volume} {252}},\
  \bibinfo {pages} {235} (\bibinfo {year} {2015})}\BibitemShut {NoStop}%
\bibitem [{\citenamefont {Rocca}\ \emph {et~al.}(2012)\citenamefont {Rocca},
  \citenamefont {Ping}, \citenamefont {Gebauer},\ and\ \citenamefont
  {Galli}}]{rocca2012solution}%
  \BibitemOpen
  \bibfield  {author} {\bibinfo {author} {\bibfnamefont {D.}~\bibnamefont
  {Rocca}}, \bibinfo {author} {\bibfnamefont {Y.}~\bibnamefont {Ping}},
  \bibinfo {author} {\bibfnamefont {R.}~\bibnamefont {Gebauer}},\ and\ \bibinfo
  {author} {\bibfnamefont {G.}~\bibnamefont {Galli}},\ }\bibfield  {title}
  {\bibinfo {title} {Solution of the bethe-salpeter equation without empty
  electronic states: Application to the absorption spectra of bulk systems},\
  }\href@noop {} {\bibfield  {journal} {\bibinfo  {journal} {Physical Review
  B}\ }\textbf {\bibinfo {volume} {85}},\ \bibinfo {pages} {045116} (\bibinfo
  {year} {2012})}\BibitemShut {NoStop}%
\bibitem [{\citenamefont {Leng}\ \emph {et~al.}(2016)\citenamefont {Leng},
  \citenamefont {Jin}, \citenamefont {Wei},\ and\ \citenamefont
  {Ma}}]{leng2016gw}%
  \BibitemOpen
  \bibfield  {author} {\bibinfo {author} {\bibfnamefont {X.}~\bibnamefont
  {Leng}}, \bibinfo {author} {\bibfnamefont {F.}~\bibnamefont {Jin}}, \bibinfo
  {author} {\bibfnamefont {M.}~\bibnamefont {Wei}},\ and\ \bibinfo {author}
  {\bibfnamefont {Y.}~\bibnamefont {Ma}},\ }\bibfield  {title} {\bibinfo
  {title} {Gw method and bethe--salpeter equation for calculating electronic
  excitations},\ }\href@noop {} {\bibfield  {journal} {\bibinfo  {journal}
  {Wiley Interdisciplinary Reviews: Computational Molecular Science}\ }\textbf
  {\bibinfo {volume} {6}},\ \bibinfo {pages} {532} (\bibinfo {year}
  {2016})}\BibitemShut {NoStop}%
\bibitem [{\citenamefont {Vlcek}(2019)}]{vlcek2019stochastic}%
  \BibitemOpen
  \bibfield  {author} {\bibinfo {author} {\bibfnamefont {V.}~\bibnamefont
  {Vlcek}},\ }\bibfield  {title} {\bibinfo {title} {Stochastic vertex
  corrections: Linear scaling methods for accurate quasiparticle energies},\
  }\href@noop {} {\bibfield  {journal} {\bibinfo  {journal} {J. Chem. Theory
  Comput.}\ }\textbf {\bibinfo {volume} {15}},\ \bibinfo {pages} {6254}
  (\bibinfo {year} {2019})}\BibitemShut {NoStop}%
\bibitem [{\citenamefont {Mejuto-Zaera}\ and\ \citenamefont
  {Vl\ifmmode~\check{c}\else \v{c}\fi{}ek}(2022)}]{mejuto2022}%
  \BibitemOpen
  \bibfield  {author} {\bibinfo {author} {\bibfnamefont {C.}~\bibnamefont
  {Mejuto-Zaera}}\ and\ \bibinfo {author} {\bibfnamefont {V.~c.~v.}\
  \bibnamefont {Vl\ifmmode~\check{c}\else \v{c}\fi{}ek}},\ }\bibfield  {title}
  {\bibinfo {title} {Self-consistency in $gw\mathrm{\ensuremath{\Gamma}}$
  formalism leading to quasiparticle-quasiparticle couplings},\ }\href
  {https://doi.org/10.1103/PhysRevB.106.165129} {\bibfield  {journal} {\bibinfo
   {journal} {Phys. Rev. B}\ }\textbf {\bibinfo {volume} {106}},\ \bibinfo
  {pages} {165129} (\bibinfo {year} {2022})}\BibitemShut {NoStop}%
\bibitem [{\citenamefont {Hedin}(1965)}]{hedin1965new}%
  \BibitemOpen
  \bibfield  {author} {\bibinfo {author} {\bibfnamefont {L.}~\bibnamefont
  {Hedin}},\ }\bibfield  {title} {\bibinfo {title} {New method for calculating
  the one-particle green's function with application to the electron-gas
  problem},\ }\href@noop {} {\bibfield  {journal} {\bibinfo  {journal}
  {Physical Review}\ }\textbf {\bibinfo {volume} {139}},\ \bibinfo {pages}
  {A796} (\bibinfo {year} {1965})}\BibitemShut {NoStop}%
\bibitem [{\citenamefont {Ranjbar}\ \emph {et~al.}(2011)\citenamefont
  {Ranjbar}, \citenamefont {Babamoradi}, \citenamefont {Saani}, \citenamefont
  {Vesaghi}, \citenamefont {Esfarjani},\ and\ \citenamefont
  {Kawazoe}}]{ranjbar2011many}%
  \BibitemOpen
  \bibfield  {author} {\bibinfo {author} {\bibfnamefont {A.}~\bibnamefont
  {Ranjbar}}, \bibinfo {author} {\bibfnamefont {M.}~\bibnamefont {Babamoradi}},
  \bibinfo {author} {\bibfnamefont {M.~H.}\ \bibnamefont {Saani}}, \bibinfo
  {author} {\bibfnamefont {M.~A.}\ \bibnamefont {Vesaghi}}, \bibinfo {author}
  {\bibfnamefont {K.}~\bibnamefont {Esfarjani}},\ and\ \bibinfo {author}
  {\bibfnamefont {Y.}~\bibnamefont {Kawazoe}},\ }\bibfield  {title} {\bibinfo
  {title} {Many-electron states of nitrogen-vacancy centers in diamond and spin
  density calculations},\ }\href@noop {} {\bibfield  {journal} {\bibinfo
  {journal} {Physical Review B}\ }\textbf {\bibinfo {volume} {84}},\ \bibinfo
  {pages} {165212} (\bibinfo {year} {2011})}\BibitemShut {NoStop}%
\bibitem [{\citenamefont {Babamoradi}\ \emph {et~al.}(2011)\citenamefont
  {Babamoradi}, \citenamefont {Heidari~Saani}, \citenamefont {Ranjbar},
  \citenamefont {Vesaghi},\ and\ \citenamefont
  {Kawazoe}}]{babamoradi2011effect}%
  \BibitemOpen
  \bibfield  {author} {\bibinfo {author} {\bibfnamefont {M.}~\bibnamefont
  {Babamoradi}}, \bibinfo {author} {\bibfnamefont {M.}~\bibnamefont
  {Heidari~Saani}}, \bibinfo {author} {\bibfnamefont {A.}~\bibnamefont
  {Ranjbar}}, \bibinfo {author} {\bibfnamefont {M.}~\bibnamefont {Vesaghi}},\
  and\ \bibinfo {author} {\bibfnamefont {Y.}~\bibnamefont {Kawazoe}},\
  }\bibfield  {title} {\bibinfo {title} {Effect of lattice relaxation on spin
  density of nitrogen-vacancy centers in diamond and oscillator strength
  calculations},\ }\href@noop {} {\bibfield  {journal} {\bibinfo  {journal}
  {The European Physical Journal B}\ }\textbf {\bibinfo {volume} {84}},\
  \bibinfo {pages} {1} (\bibinfo {year} {2011})}\BibitemShut {NoStop}%
\bibitem [{\citenamefont {Romanova}\ and\ \citenamefont
  {Vl{\v{c}}ek}(2020)}]{romanova2020decomposition}%
  \BibitemOpen
  \bibfield  {author} {\bibinfo {author} {\bibfnamefont {M.}~\bibnamefont
  {Romanova}}\ and\ \bibinfo {author} {\bibfnamefont {V.}~\bibnamefont
  {Vl{\v{c}}ek}},\ }\bibfield  {title} {\bibinfo {title} {Decomposition and
  embedding in the stochastic {GW} self-energy},\ }\href@noop {} {\bibfield
  {journal} {\bibinfo  {journal} {J. Chem. Phys.}\ }\textbf {\bibinfo {volume}
  {153}},\ \bibinfo {pages} {134103} (\bibinfo {year} {2020})}\BibitemShut
  {NoStop}%
\bibitem [{\citenamefont {Faleev}\ \emph {et~al.}(2004)\citenamefont {Faleev},
  \citenamefont {Van~Schilfgaarde},\ and\ \citenamefont
  {Kotani}}]{faleev2004all}%
  \BibitemOpen
  \bibfield  {author} {\bibinfo {author} {\bibfnamefont {S.~V.}\ \bibnamefont
  {Faleev}}, \bibinfo {author} {\bibfnamefont {M.}~\bibnamefont
  {Van~Schilfgaarde}},\ and\ \bibinfo {author} {\bibfnamefont {T.}~\bibnamefont
  {Kotani}},\ }\bibfield  {title} {\bibinfo {title} {All-electron
  self-consistent {GW} approximation: Application to {S}i, {M}n{O}, and
  {N}i{O}},\ }\href@noop {} {\bibfield  {journal} {\bibinfo  {journal} {Phys.
  Rev. Lett.}\ }\textbf {\bibinfo {volume} {93}},\ \bibinfo {pages} {126406}
  (\bibinfo {year} {2004})}\BibitemShut {NoStop}%
\bibitem [{\citenamefont {Bruneval}\ \emph {et~al.}(2006)\citenamefont
  {Bruneval}, \citenamefont {Vast},\ and\ \citenamefont
  {Reining}}]{bruneval2006effect}%
  \BibitemOpen
  \bibfield  {author} {\bibinfo {author} {\bibfnamefont {F.}~\bibnamefont
  {Bruneval}}, \bibinfo {author} {\bibfnamefont {N.}~\bibnamefont {Vast}},\
  and\ \bibinfo {author} {\bibfnamefont {L.}~\bibnamefont {Reining}},\
  }\bibfield  {title} {\bibinfo {title} {Effect of self-consistency on
  quasiparticles in solids},\ }\href@noop {} {\bibfield  {journal} {\bibinfo
  {journal} {Phys. Rev. B}\ }\textbf {\bibinfo {volume} {74}},\ \bibinfo
  {pages} {045102} (\bibinfo {year} {2006})}\BibitemShut {NoStop}%
\bibitem [{\citenamefont {Onida}\ \emph {et~al.}(2002)\citenamefont {Onida},
  \citenamefont {Reining},\ and\ \citenamefont {Rubio}}]{onida2002electronic}%
  \BibitemOpen
  \bibfield  {author} {\bibinfo {author} {\bibfnamefont {G.}~\bibnamefont
  {Onida}}, \bibinfo {author} {\bibfnamefont {L.}~\bibnamefont {Reining}},\
  and\ \bibinfo {author} {\bibfnamefont {A.}~\bibnamefont {Rubio}},\ }\bibfield
   {title} {\bibinfo {title} {Electronic excitations: density-functional versus
  many-body green’s-function approaches},\ }\href@noop {} {\bibfield
  {journal} {\bibinfo  {journal} {Reviews of modern physics}\ }\textbf
  {\bibinfo {volume} {74}},\ \bibinfo {pages} {601} (\bibinfo {year}
  {2002})}\BibitemShut {NoStop}%
\bibitem [{\citenamefont {Neuhauser}\ \emph {et~al.}(2014)\citenamefont
  {Neuhauser}, \citenamefont {Gao}, \citenamefont {Arntsen}, \citenamefont
  {Karshenas}, \citenamefont {Rabani},\ and\ \citenamefont
  {Baer}}]{neuhauser2014breaking}%
  \BibitemOpen
  \bibfield  {author} {\bibinfo {author} {\bibfnamefont {D.}~\bibnamefont
  {Neuhauser}}, \bibinfo {author} {\bibfnamefont {Y.}~\bibnamefont {Gao}},
  \bibinfo {author} {\bibfnamefont {C.}~\bibnamefont {Arntsen}}, \bibinfo
  {author} {\bibfnamefont {C.}~\bibnamefont {Karshenas}}, \bibinfo {author}
  {\bibfnamefont {E.}~\bibnamefont {Rabani}},\ and\ \bibinfo {author}
  {\bibfnamefont {R.}~\bibnamefont {Baer}},\ }\bibfield  {title} {\bibinfo
  {title} {{Breaking the Theoretical Scaling Limit for Predicting Quasiparticle
  Energies: The Stochastic GW Approach}},\ }\href@noop {} {\bibfield  {journal}
  {\bibinfo  {journal} {Phys. Rev. Lett.}\ }\textbf {\bibinfo {volume} {113}},\
  \bibinfo {pages} {076402} (\bibinfo {year} {2014})}\BibitemShut {NoStop}%
\bibitem [{\citenamefont {Vl\v{c}ek}\ \emph {et~al.}(2017)\citenamefont
  {Vl\v{c}ek}, \citenamefont {Rabani}, \citenamefont {Neuhauser},\ and\
  \citenamefont {Baer}}]{vlcek2017stochastic}%
  \BibitemOpen
  \bibfield  {author} {\bibinfo {author} {\bibfnamefont {V.}~\bibnamefont
  {Vl\v{c}ek}}, \bibinfo {author} {\bibfnamefont {E.}~\bibnamefont {Rabani}},
  \bibinfo {author} {\bibfnamefont {D.}~\bibnamefont {Neuhauser}},\ and\
  \bibinfo {author} {\bibfnamefont {R.}~\bibnamefont {Baer}},\ }\bibfield
  {title} {\bibinfo {title} {Stochastic {GW} calculations for molecules},\
  }\href@noop {} {\bibfield  {journal} {\bibinfo  {journal} {J. Chem. Theory
  Comput.}\ }\textbf {\bibinfo {volume} {13}},\ \bibinfo {pages} {4997}
  (\bibinfo {year} {2017})}\BibitemShut {NoStop}%
\bibitem [{\citenamefont {Giannozzi}\ \emph {et~al.}(2017)\citenamefont
  {Giannozzi} \emph {et~al.}}]{QE2017}%
  \BibitemOpen
  \bibfield  {author} {\bibinfo {author} {\bibfnamefont {P.}~\bibnamefont
  {Giannozzi}} \emph {et~al.},\ }\bibfield  {title} {\bibinfo {title} {Advanced
  capabilities for materials modelling with quantum espresso},\ }\href@noop {}
  {\bibfield  {journal} {\bibinfo  {journal} {J. Condens. Matter Phys.}\
  }\textbf {\bibinfo {volume} {29}},\ \bibinfo {pages} {465901} (\bibinfo
  {year} {2017})}\BibitemShut {NoStop}%
\bibitem [{\citenamefont {Weng}\ and\ \citenamefont
  {Vlček}(2021)}]{gwen_wannier}%
  \BibitemOpen
  \bibfield  {author} {\bibinfo {author} {\bibfnamefont {G.}~\bibnamefont
  {Weng}}\ and\ \bibinfo {author} {\bibfnamefont {V.}~\bibnamefont {Vlček}},\
  }\bibfield  {title} {\bibinfo {title} {Efficient treatment of molecular
  excitations in the liquid phase environment via stochastic many-body
  theory},\ }\href {https://doi.org/10.1063/5.0058410} {\bibfield  {journal}
  {\bibinfo  {journal} {The Journal of Chemical Physics}\ }\textbf {\bibinfo
  {volume} {155}},\ \bibinfo {pages} {054104} (\bibinfo {year} {2021})},\
  \Eprint {https://arxiv.org/abs/https://doi.org/10.1063/5.0058410}
  {https://doi.org/10.1063/5.0058410} \BibitemShut {NoStop}%
\bibitem [{\citenamefont {Weng}\ \emph {et~al.}(2022)\citenamefont {Weng},
  \citenamefont {Romanova}, \citenamefont {Apelian}, \citenamefont {Song},\
  and\ \citenamefont {Vlček}}]{weng2022reduced}%
  \BibitemOpen
  \bibfield  {author} {\bibinfo {author} {\bibfnamefont {G.}~\bibnamefont
  {Weng}}, \bibinfo {author} {\bibfnamefont {M.}~\bibnamefont {Romanova}},
  \bibinfo {author} {\bibfnamefont {A.}~\bibnamefont {Apelian}}, \bibinfo
  {author} {\bibfnamefont {H.}~\bibnamefont {Song}},\ and\ \bibinfo {author}
  {\bibfnamefont {V.}~\bibnamefont {Vlček}},\ }\bibfield  {title} {\bibinfo
  {title} {Reduced scaling of optimal regional orbital localization via
  sequential exhaustion of the single-particle space},\ }\href@noop {}
  {\bibfield  {journal} {\bibinfo  {journal} {Journal of Chemical Theory and
  Computation}\ } (\bibinfo {year} {2022})}\BibitemShut {NoStop}%
\bibitem [{\citenamefont {Choi}\ \emph {et~al.}(2012)\citenamefont {Choi},
  \citenamefont {Jain},\ and\ \citenamefont {Louie}}]{choi2012mechanism}%
  \BibitemOpen
  \bibfield  {author} {\bibinfo {author} {\bibfnamefont {S.}~\bibnamefont
  {Choi}}, \bibinfo {author} {\bibfnamefont {M.}~\bibnamefont {Jain}},\ and\
  \bibinfo {author} {\bibfnamefont {S.~G.}\ \bibnamefont {Louie}},\ }\bibfield
  {title} {\bibinfo {title} {Mechanism for optical initialization of spin in
  nv- center in diamond},\ }\href@noop {} {\bibfield  {journal} {\bibinfo
  {journal} {Physical Review B}\ }\textbf {\bibinfo {volume} {86}},\ \bibinfo
  {pages} {041202} (\bibinfo {year} {2012})}\BibitemShut {NoStop}%
\bibitem [{\citenamefont {Nebel}\ and\ \citenamefont
  {Ristein}(2003)}]{nebel2003thin}%
  \BibitemOpen
  \bibfield  {author} {\bibinfo {author} {\bibfnamefont {C.}~\bibnamefont
  {Nebel}}\ and\ \bibinfo {author} {\bibfnamefont {J.}~\bibnamefont
  {Ristein}},\ }\href@noop {} {\emph {\bibinfo {title} {Thin-Film Diamond
  I:(part of the Semiconductors and Semimetals Series)}}}\ (\bibinfo
  {publisher} {Academic Press},\ \bibinfo {year} {2003})\BibitemShut {NoStop}%
\bibitem [{\citenamefont {Davies}\ and\ \citenamefont
  {Hamer}(1976)}]{davies1976optical}%
  \BibitemOpen
  \bibfield  {author} {\bibinfo {author} {\bibfnamefont {G.}~\bibnamefont
  {Davies}}\ and\ \bibinfo {author} {\bibfnamefont {M.}~\bibnamefont {Hamer}},\
  }\bibfield  {title} {\bibinfo {title} {Optical studies of the 1.945 ev
  vibronic band in diamond},\ }\href@noop {} {\bibfield  {journal} {\bibinfo
  {journal} {Proceedings of the Royal Society of London. A. Mathematical and
  Physical Sciences}\ }\textbf {\bibinfo {volume} {348}},\ \bibinfo {pages}
  {285} (\bibinfo {year} {1976})}\BibitemShut {NoStop}%
\bibitem [{\citenamefont {Towns}\ \emph {et~al.}(2014)\citenamefont {Towns}
  \emph {et~al.}}]{Towns_2014}%
  \BibitemOpen
  \bibfield  {author} {\bibinfo {author} {\bibfnamefont {J.}~\bibnamefont
  {Towns}} \emph {et~al.},\ }\bibfield  {title} {\bibinfo {title} {Xsede:
  Accelerating scientific discovery},\ }\href
  {https://doi.org/10.1109/MCSE.2014.80} {\bibfield  {journal} {\bibinfo
  {journal} {Comput. Sci. Eng.}\ }\textbf {\bibinfo {volume} {16}},\ \bibinfo
  {pages} {62} (\bibinfo {year} {2014})}\BibitemShut {NoStop}%
\end{thebibliography}%

\end{document}